\documentclass[aps,prb]{revtex4}
\usepackage{amsmath,amssymb}
\usepackage{graphicx}
\usepackage{dcolumn}
\usepackage{bm}
\usepackage{color}
\usepackage{relsize}
\newcommand{\mbb}{\mathbb}
\newcommand{\mc}{\mathcal}

\newcommand{\tet}{\texttt}
\newcommand{\pr}{\partial}

\begin{document}
\title{Anomalous Klein paradox due to misalignment of optically-tunable elliptical dispersion for Dirac-cone dressed states and direction of incoming particles}

\author{Andrii Iurov$^{1} \footnote{E-mail contact: aiurov@mec.cuny.edu, theorist.physics@gmail.com
}$,
Liubov Zhemchuzhna$^{2,1}$,
Paula Fekete$^3$,
Godfrey Gumbs$^{2,4}$, 
and
Danhong Huang$^{5,6}$
}

\affiliation{
$^{1}$Department of Physics and Computer Science, Medgar Evers College of City University of New York, Brooklyn, NY 11225, USA\\ 
$^{2}$Department of Physics and Astronomy, Hunter College of the City University of New York, 695 Park Avenue, New York, New York 10065, USA\\ 
$^{3}$US Military Academy at West Point, 606 Thayer Road, West Point, New York 10996, USA\\
$^{4}$Donostia International Physics Center (DIPC), P de Manuel Lardizabal, 4, 20018 San Sebastian, Basque Country, Spain\\ 
$^{5}$US Air Force Research Laboratory, Space Vehicles Directorate, Kirtland Air Force Base, New Mexico 87117, USA\\ 
$^{6}$Center for High Technology Materials, University of New Mexico, 1313 Goddard SE, Albuquerque, New Mexico, 87106, USA\\
}

\date{\today}

\begin{abstract}

After having derived boundary conditions for dressed-state electrons in a dice lattice, 
we investigate the electron tunneling through a square electrostatic potential barrier in both dice lattices and graphene under a 
linearly-polarized off-resonance and high-frequency dressing field, and demonstrate the anomalous Klein paradox for a nonzero incident angle, 
resulted from the misalignment of optically-controllable elliptical dispersion 
for Dirac-cone dressed states and the direction of incoming kinetic particles in our system. 
This finite incident angle is found depending on the type of light polarization, the light-induced anisotropy in energy dispersion and the strength of electron-light coupling. 
Meanwhile, we also observe much larger off-peak transmission amplitudes in dice lattices in comparison with graphene.
We expect the theoretical results in this paper could be used for wide range of Dirac materials and applied to controlling both coherent tunneling and ballistic transport of electrons 
for constructing novel optical and electronic nano-scale switching devices.   
\end{abstract}

\maketitle

\section{Introduction} 
\label{sec1}

The $\alpha-\mc{T}_3$ model is the newest and the most promising system with zero-mass Dirac fermions\,\cite{thesis}.
Unlike graphene\,\cite{gr01, neto}, the low-energy electronic states of $\alpha-\mc{T}_3$ lattices are governed
by a $3 \times 3$ pseudospin-$1$ Hamiltonian, and meanwhile are described mathematically by pseudospin-$1$ Dirac-Weyl equation\,\cite{Dora12, Malc01, vidal1, vidal2}. 
The resulting energy dispersion is distinguished because of a completely flat band with infinite degeneracy exactly at the 
Dirac point, and also acquires a Dirac-cone structure as in graphene at the same time. These Dirac-cone bands could open an energy gap or display an anisotropy if 
an external irradiation with a proper polarization is applied to the system\,\cite{kiMain, ourpeculiar}. 
\medskip

On the atomic-structure level, the construction of an $\alpha-\mc{T}_3$ lattice can be viewed simply as a honeycomb lattice of graphene plus an additional hub atom at the center of each hexagon.
This hub atom couples to one of the $A$- or $B$-sublattice atom on the rim with its hopping coefficient equal to a fraction of that between two neighboring
sublattice-atoms on the hexagon rim sites. This ratio $\alpha$ differs from $0$, in contrast to graphene with $\alpha=0$, for a 
decoupled and non-interacting set of hub atoms or is equal to $1$ corresponding to a dice lattice in which the influence of the extra
hub atom reaches a maximum. 
\medskip

There has been a great deal of encouraging experimental evidence for the fabrication\,\cite{Dan1, Dan2, Dan5, Dan6, Dan7} of a dice or $\alpha-\mc{T}_3$ lattices
based on various atomic and electronic properties of some known materials\,\cite{R110, R115, R116, R117, R133}. One of the most well-known and widely 
discussed candidates is the three-layer arrangement of SrTiO$_3$/SrIrO$_3$/SrTiO$_3$ lattices in which each of three initial layers possesses a cubic crystal structure. 
A particularly useful and complete review on the experimentally-fabricated flat-band materials could be found in Ref.\,[\onlinecite{Add1}] and the cited references therein.   
\medskip

Novel $\alpha-\mc{T}_3$ model has already demonstrated nontrivial topology\,\cite{dey1, dey2, dey3, ber2} related to its band structure due to the 
presence of an additional flat band, and the unique topological features have been seen from many of its physical properties, 
including both optical\,\cite{opt1, ourLast} and magnetic\,\cite{tutul1, tutul2, nic1, nic2} ones. 
The most fascinating characteristics exhibited is the phase transition from a diamagnetic to a paramagnetic material under a perpendicular quantizing magnetic field 
as the $\alpha$ parameter is increased from zero.\,\cite{piech1, piech2}
Meanwhile, the electronic\,\cite{ourLor, dey1}, collective\,\cite{Malc01, ourLast,ml} and transport\,\cite{wa20, c0, our20} phenomena in $\alpha-\mc{T}_3$ are also found unique and remarkable. 
Especially, $\alpha-\mc{T}_3$ materials allow for regular Klein paradox,\,\cite{alphaKlein, alphaDice} i.e., unimpeded tunneling for Dirac electrons normally incident on a sharp potential barrier,\,\cite{fanwar} just as it was observed in graphene\,\cite{kacm, k223, pet1, pet2, pet3} earlier.
Interestingly, such a complete transmission is independent of the barrier height and width. 
\medskip 

Physically, the electronic states and their properties in all these newly discovered low-dimensional materials could be modified effectively and even 
tuned finely by employing the so-called Floquet engineering, i.e., applying an off-resonance and high-frequency dressing field with various polarizations.
The practical use of such a semi-classical dressing approach with a non-ionizing but intensive laser field has only become possible over the last several years due to significant
progress made in microwave, laser and teraherze technologies.  
The modification of electronic properties based on the external irradiation has been addressed theoretically in an extensive way with the help from 
Floquet theory\,\cite{p1, p2, p3} for periodically driven quantum systems\,\cite{prx1}, covering an extremely wide range of two-dimensional materials\,\cite{Tor1, Mor1}, 
such as, graphene\,\cite{kiMain, kisrep}, silicene and transitional-metal dischalcogenides\,\cite{kibisAll, kn2}
phosphorenes\,\cite{ourJAP2017}, purely quantum-spin systems\,\cite{kn1} 
and on the surfaces of three-dimensional topological insulators\,\cite{kn3, kn4}. 
\medskip

The effect of an applied irradiation onto a two-dimensional material depends greatly on its polarization. 
Circularly-polarized light leads to opening a bandgap between the valence and conduction bands of an isotropic energy spectrum\,\cite{kiMain} as well as the suppression 
of the Klein tunneling\,\cite{oura, kn5}. This opened energy gap has an important effect on the collective electronic properties of $\alpha-\mc{T}_3$ lattices\,\cite{ourcontrolling, ourC18}. 
Linearly-polarized irradiation, on the other hand, creates an anisotropy in the Dirac dispersion\,\cite{kisrep} or modifies the existing anisotropy 
within the phosphoene band structure\,\cite{LiuM, ourJAP2017}, 
which is equivalent to applying the most general elliptically-polarized dressing field with combined anisotropies from both the material band structure and the external light-field polarization.
\medskip 

The rest of the paper is organized as follows. First, we will provide an alternative derivation in Sec.\,\ref{sec2} for the dressed electronic states in a dice
lattice using a rather straightforward Floquet-Magnus perturbative expansion for an off-resonance and high-frequency periodic dressing field. 
The obtained  electronic states in Sec.\,\ref{sec2} will be compared with our early derived results\,\cite{ourpeculiar} based on a rigorous analytical solution for $\mbox{\boldmath$k$}=0$ 
followed by an expansion with respect to the complete set at $\mbox{\boldmath$k$}=0$ for other $\mbox{\boldmath$k$}$ vectors. 
The corresponding derivation of such dressed states for arbitrary direction of linear polarization is briefly discussed in Appendix\ \ref{apb}.
Equipped with the obtained dressed state of electrons, we further study the electron-tunneling dynamics through a square-barrier potential 
under an external linearly-polarized dressing field for both graphene and a dice lattice. 
We demonstrate in Appendix\ \ref{apd} that the boundary conditions for a dice lattice should be modified substantially if the direction of light polarization and the direction of a head-on incidence is misaligned, which is linked to the so-called anomalous Klein paradox. 
We find the expressions in Section\ \ref{sec3} for electron tunneling in a dice lattice or graphene with an anisotropic Dirac cone and obtain the results for anomalous Klein paradox. 
In Sec.\,\ref{sec4}, we analyze and discuss the properties of obtained numerical results for electron transmission and reflection in both irradiated graphene and dice lattice, 
and draw the final conclusions as well as remarks in Sec.\,\ref{sec5}.          

\section{Electron-Dressed States Under Linearly-Polarized Irradiation} 
\label{sec2}

In this section, we present an alternative, much simplified, derivation of the electron dressed states in the presence of external 
linearly-polarized irradiation. Using the Floquet-Magnus perturbation expansion designed for the off-resonance dressing field with the 
frequency $\hbar\omega \gg \mc{E}_0$ the characteristic energy of electrons, we obtain the quasiparticle energy dispersions 
and the closed-form analytical expression for dressed electron wave functions. 
Even though this paper focuses on two opposite limits of graphene $\alpha = 0$ and a dice lattice $\alpha = 1$,
we will still present some relevant discussions on properties pertaining to the general case of an $\alpha-\mc{T}_3$ model. 
\medskip

We begin with the low-energy Hamiltonian for $\alpha-\mc{T}_3$ materials under applied {\it linearly-polarized} radiation with 
a vector potential $\mbox{\boldmath$A$}^{(L)}(t)$ and an electrostatic barrier potential\,\cite{alphaDice} $V(x)=V_B\,\Theta(x)\,\Theta(W_B-x)$, i.e.,

\begin{equation}
\label{HamG}
\hat{\mbb{H}}_0^\tau (\phi \, \vert \, x, y) = v_F \, \hat{\mbox{\boldmath$S$}}(\phi) \cdot \left\{ -i \hbar\mbox{\boldmath$\nabla$}_{\{\tau\}} - e\mbox{\boldmath$A$}^{(L)}(t) \right\} + V(x) \ ,
\end{equation}
where $V_B$ and $W_B$ are the strength and width of the barrier potential, 
the two $\phi$-dependent matrices $\hat{\mbox{\boldmath$S$}}(\phi)= \left\{  \hat{S}_x(\phi),\,\hat{S}_y(\phi) \right\}$ are defined 
in Appendix\ \ref{apa}, $\mbox{\boldmath$\nabla$}_{\{\tau\}}= \{ \tau \pr/\pr x,\,\pr/\pr y \}$ is the two-dimensional gradient operator depending on the valley index $\tau = \pm 1$. 
\medskip

\begin{figure} 
\centering
\includegraphics[width=0.45\textwidth]{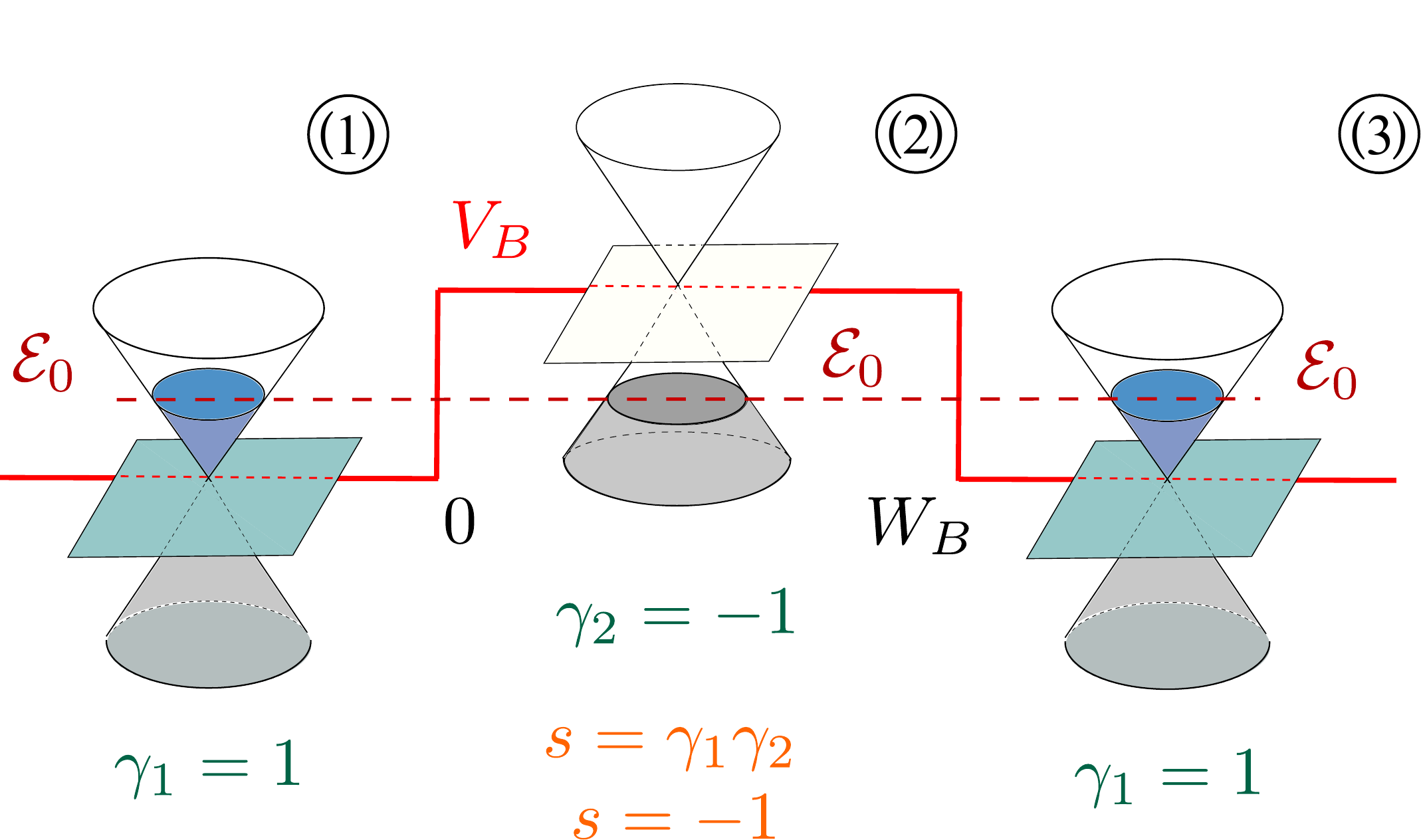}
\caption{(Color online)  Schematics for an incident electron with kinetic energy $\varepsilon_0$ tunneling through a rectangular potential barrier $V(x)=V_B\,\Theta(x)\Theta(W_B - x)$.
We have chosen the barrier-height $V_B$ such that $0 <\varepsilon_0 < V_B$ and the electron-hole-electron transition occurs between two edges,
$x=0$ and $x=W_B$, of the barrier region-$2$, which is equivalent to $n$-$p$-$n$ multi-junctions. Here, $\gamma=+1$ ($\gamma=-1$) corresponds to the Fermi energy sitting within the 
upper (lower) Dirac cone, and $\varepsilon_0$ is the incident kinetic energy of electrons.
Furthermore, the energy unit for $\varepsilon_0$ and $V_B$ is the Fermi energy $E_F^{(0)}$, while the length unit for $W_B$ is $1/k_F^{(0)}$, 
where $k_F^{(0)}=\sqrt{\pi n_0}$ is the Fermi wave number and $E_F^{(0)}=\hbar v_Fk_F^{(0)}$ with $v_F$ and $n_0$ as the Fermi velocity and the areal doping density, respectively.}
\label{FIG:1}
\end{figure}

The potential $V(x)$ in Eq.\,\eqref{HamG} only relies on the position $x$ but not $y$. Here, $V(x)$ is assumed a piecewise-constant profile 
as commonly employed for studying Klein tunneling\,\cite{kacm, alphaDice, alphaKlein}. 
Moreover, $V(x)$ brings into two boundary conditions at its edges while keeps a translational symmetry along the $y$ direction of the system. 
On the other hand, the physics characteristics of carriers, i.e., electrons or holes, in the barrier region is determined by the sign of $\varepsilon_0-V_0$, 
where $\varepsilon_0$ represents the energy of incoming electrons. Since we consider a positive barrier with $V_B>0$, there exists only one transition in the barrier region, 
i.e., electron $\rightarrow$ hole, under the condition of $V_B >\varepsilon_0$, as shown in Fig.\,\ref{FIG:1}. 
\medskip

By taking $V(x)=0$ in Eq.\,\eqref{HamG}, due to the presence of translational symmetry in the system, we acquire the simple relation, i.e., $ \{\pr/ \pr x, \pr/ \pr y \} 
\rightarrow i \, \{k_x, k_y\}$, for the wave function $\Psi(x,y) \backsim \tet{e}^{i k_x \, x} \, 
\tet{e}^{i k_y \, y}$, and the first term of the Hamiltonian in Eq.\,\eqref{HamG} becomes

\begin{equation}
\label{H0}
\hat{\mc{H}}_{\alpha} (\mbox{\boldmath$k$} \, \vert \, \tau, \phi) = \hbar v_F \,
\left[
\begin{array}{ccc}
0 & k^\tau_- \, \cos \phi &  0 \\
k^\tau_+ \, \cos \phi & 0 & k^\tau_- \, \sin \phi \\
0 & k^\tau_+ \, \sin \phi & 0
\end{array}
\right]
+ \text{h.c.}\ ,  
\end{equation}
where $k^\tau_\pm = \tau k_x \pm i k_y$ depending on the valley index $\tau = \pm 1$, the geometry phase $\phi$ is related to the ratio of the hopping 
amplitudes $\alpha$ by $\phi = \tan^{-1} \alpha$ for $0\leq\phi\leq\pi/4$, (later we will only consider a dice lattice with $\alpha = 1$ and 
$\phi = \pi/4$) and $+\text{h.c.}$ means adding the Hermitian conjugate of the first term. 
From now on, we will only consider a dice lattice with $\alpha = 1$ or $\phi = \pi/4$. 
\medskip 

Particularly, for the case of a dice lattice with $\phi = \pi/4$, the Hamiltonian in Eq.\,\eqref{H0} for $V(x)=0$ is simplified as 

\begin{equation}
\label{H0D}
\hat{\mc{H}}_{1} (\mbox{\boldmath$k$} \, \vert \, \tau) = \frac{\hbar v_F}{\sqrt{2}} \, \left[
\begin{array}{ccc}
0 & k^\tau_{-} & 0 \\
k^\tau_{+} & 0 & k^\tau_{-} \\
0 & k^\tau_{+} & 0
\end{array}
\right] = 
\sum \limits_{s = \pm} \hat{\Sigma}_\alpha^{\,(1)} \, k^\tau_{s} \ ,
\end{equation}
where $\hat{\Sigma}_{\pm 1}^{\,(1)} = \hat{\Sigma}_{x}^{\,(1)} \pm i\,\hat{\Sigma}_{y}^{\,(1)}$, which are defined based
on the spin-$1$ matrices, are derived and explained in Appendix\ \ref{apa}.
\medskip 

Our primary goal in his paper is to find the electron dressed states under a linearly-polarized dressing field. Here, we assume the polarization 
of the dressing field lies in the $x$ direction, while the general case with an arbitrary polarization direction is
discussed in Appendix\ \ref{apb}. Under this assumption,  the vector potential takes the form 

\begin{equation}
\label{linA}
\mbox{\boldmath$A$}^{(L)}(t) = 
\left[  \begin{array}{c}
          A^{(L)}_x (t) \\
          A^{(L)}_y (t)
        \end{array}
\right] = \frac{E_0}{\omega} \left[
\begin{array}{c}
\cos \beta\\
0
\end{array}
\right] \, \cos (\omega t ) \  .
\end{equation}
This is one limiting case for the most general elliptically-polarized light, and the opposite limit with two equal components of the vector
potential corresponds to the circularly-polarized light. Under the linearly-polarized irradiation, the wave vector $\mbox{\boldmath$k$}$ in the Hamiltonian 
in Eq.\,\eqref{H0D} is modified accordingly based on the 
canonical substitution, i.e., $k_{x,y} \rightarrow k_{x,y} - e\,A_{x,y}/\hbar$.
\medskip

Since the Hamiltonian in Eq.\eqref{H0D} is linear in $k_{x,y}$, in the presence of $\mbox{\boldmath$A$}^{(L)}(t)$ it only acquires an additional {\it interaction} term, yielding

\begin{equation}
\label{Tlinham}
\hat{\mc{H}}_{1} (\mbox{\boldmath$k$}\, \vert \, \tau) \Longrightarrow \hat{\mbb{H}}^{(L)}(\mbox{\boldmath$k$}, t) = 
\hat{\mc{H}}_{1} (\mbox{\boldmath$k$}\, \vert \, \tau) + \hat{\mc{H}}_A^{(L)}(t) \ , 
\end{equation}
where the $\mbox{\boldmath$k$}$ independent interaction term is

\begin{equation}
 \label{HAL}
 \hat{\mc{H}}_A^{(L)}(t) = - \frac{\tau \, c_0}{\sqrt{2}}  \cos (\omega t) \left[
 \begin{array}{ccc}
  0 & 1 & 0 \\
  1  & 0 & 1 \\
  0 & 1 & 0 
 \end{array}
 \right] \ .
\end{equation}
The optical-coupling constant $c_0 = ev_F E_0/\omega$ remains the same for all types of the light polarizations which implies that its
effect on the energy dispersions has a similar magnitude but different features. In fact, the time-dependent second term in Eq.\,\eqref{HAL} 
is the same for all matrix elements of $\hat{\mc{H}}_{1} (\mbox{\boldmath$k$}\, \vert \, \tau)$, which is unique for the linear type of the imposed light polarization. 
\medskip

In this paper, we apply the Floquet-Magnus perturbation approach to the Hamiltonian in Eq.\,\eqref{Tlinham} for a high-frequency 
off-resonance dressing field. For this purpose, we first rewrite the time-dependent second term in Eq.\,\eqref{HAL} into 

\begin{equation}
\hat{\mc{H}}_A^{(L)}(t)  = \hat{\mbb{O}}_1(c_0, \tau) \, \tet{e}^{i \omega t} + 
\hat{\mbb{O}}_1^{\dagger}(c_0, \tau) \, \tet{e}^{-i \omega t} \ , 
\end{equation}
where the time-independent operator $\hat{\mbb{O}}_1(c_0, \tau)$ is defined as

\begin{equation}
\label{opO}
\hat{\mbb{O}}_1(c_0, \tau) = - \frac{\tau \, c_0}{2 \sqrt{2}} \,
\left[
\begin{array}{ccc}
0 & 1 & 0 \\
1 & 0 & 1 \\
0 & 1 & 0
\end{array}
\right] =  - \frac{\tau \, c_0}{2 \sqrt{2}} \, \left(\hat{\Sigma}_+^{\,(1)} + \hat{\Sigma}_-^{\,(1)}\right) \ . 
\end{equation}
Next, by employing the high-frequency Floquet-Magnus expansion technique, the time-independent effective part of the total Hamiltonian in Eq.\eqref{Tlinham} becomes

\begin{eqnarray}
\nonumber
\hat{\mc{H}}_{\text{eff}}^{\,(L)}(\mbox{\boldmath$k$}\, \vert \, \tau)& =&  \hat{\mc{H}}_{1}(\mbox{\boldmath$k$}\, \vert \, \tau) + \frac{1}{\hbar \omega} \, \left[ \, \hat{\mbb{O}}_1(c_0, \tau) , \, \hat{\mbb{O}}_1^{\dagger}(c_0, \tau)  \,\right]\\
\label{Tmexp}
& +& \frac{1}{2 (\hbar \omega)^2} \left\{
 \left[ \left[
 \, \hat{\mbb{O}}_1(c_0, \tau), \, \hat{\mc{H}}_{\, 1} (\mbox{\boldmath$k$}\, \vert \, \tau) \, 
 \right], \, 
 \hat{\mbb{O}}_1^{\dagger}(c_0, \tau) \,  
 \right]
 \,\, + \,\, h.c.
 \right\} \,\, + \cdots \ .
\end{eqnarray}
Here, the first term in Eq.\,\eqref{Tmexp} is just the non-perturbed Hamiltonian in Eq.\,\eqref{H0D} in the absence of irradiation, 
and the following one $\left[ \, \hat{\mbb{O}}_1(c_0, \tau), \, \hat{\mbb{O}}_1^{\dagger}(c_0, \tau) \right]$ 
is zero since $\hat{\mbb{O}}_1(c_0, \tau)$ is Hermitian. This is true only for linearly-polarized light 
in contrast to all other polarizations and cases with a finite bandgap. 
The remaining term in Eq.\,\eqref{Tmexp} is written as $\hat{\mbb{T}}_2(\lambda_0 \, \vert \, k, \theta_{\bf k})$ and calculated as 

\begin{equation}
\label{t20}
\hat{\mbb{T}}_2(\lambda_0 \, \vert \, k, \theta_{\bf k}) = i \frac{\lambda_0^2}{4 \sqrt{2}} \,\hbar v_Fk_y 
\left[
\begin{array}{ccc}
0 & 1 & 0 \\
-1 & 0 & 1 \\
0 & -1 & 0
\end{array}
\right] =  - \frac{\lambda_0^2}{4}\,\hbar v_F k_y  \, \hat{\Sigma}_y^{\,(1)} \ ,
\end{equation}
where $\lambda_0=c_0/\hbar\omega$ is a dimensionless interaction parameter.
\medskip

Once the full Hamiltonian for dressed-state electrons is obtained, we are able to solve the corresponding eigenvalue equation 
and find the dispersion for all energy bands. In addition to the flat-band $\varepsilon^{\gamma = 0} (\lambda_0, \mbox{\boldmath$k$})=0$, we also get other two bands

\begin{equation}
\label{en}
\varepsilon_{\, 1}^{\gamma = \pm 1} (\lambda_0, \mbox{\boldmath$k$}) =
\gamma\,\hbar v_F\,\sqrt{k_x^2 + \left( 1 - \frac{\lambda_0^2}{4} \right)^2k_y^2 }
\end{equation}
for the valence ($\gamma=-1$) and conduction ($\gamma=+1$) dressed-state electrons. Here, we introduce an anisotropic-dispersion factor $a_1(\lambda_0)$ defined by  
$\varepsilon_{\, 1}^{\gamma = \pm 1} (\lambda_0, \mbox{\boldmath$k$})=\pm\,\hbar v_F\sqrt{k_x^2+a^2_1(\lambda_0)\,k^2_y}$ and find

\begin{equation}
\label{adice}
a_1(\lambda_0) =1 - \frac{\lambda_0^2}{4}\ .
\end{equation}
Equation\ \eqref{adice} agrees with the previous result\,\cite{ourpeculiar} in the limit of $\phi \Longrightarrow \pi/4$ (or $\alpha \Longrightarrow 1$) 
for the energy dispersion of an irradiated $\alpha-\mc{T}_3$ lattice, given by

\begin{eqnarray}
\label{linD}
 && \varepsilon_{\, \alpha}^{\gamma = \pm 1} (\lambda_0, \mbox{\boldmath$k$}) = 0 \,\,\, \text{and} \\
 \nonumber 
 && \varepsilon_{\, \alpha}^{\gamma = \pm 1} (\lambda_0, \mbox{\boldmath$k$}) = \pm \hbar v_F k \, 
 \sqrt{\mc{F}(\theta_{\bf k} \, \vert \, \phi, \lambda_0)}\ ,
\end{eqnarray}
where 

\begin{equation}
\label{TA}
\mc{F}(\theta_{\bf k} \, \vert \, \phi, \lambda_0) = 
 \cos^2 \theta_{\bf k} + \left[
 J^2_0 (2 \lambda_0) \, \cos^2(2 \phi) + 
 J^2_0 (\lambda_0) \, \sin^2(2 \phi)
 \right] \,\sin^2 \theta_{\bf k}\ ,
\end{equation} 
$J_0 (x)$ is the zeroth-order Bessel function of the first kind, and the anisotropic factor of the dispersion is calculated as

\begin{equation}
\label{aphi}
a_\alpha(\lambda_0) = 1 - \frac{\lambda_0^2}{8} \left[ 5 + 3 \cos (4 \phi) \,\right] \ .
\end{equation}
Clearly, Eqs.\,\eqref{linD} and \eqref{aphi} for $\phi=\pi/4$ matches exactly Eqs.\,\eqref{en} and \eqref{adice} for a dice lattice. 
\medskip 

\begin{figure} 
\centering
\includegraphics[width=0.9\textwidth]{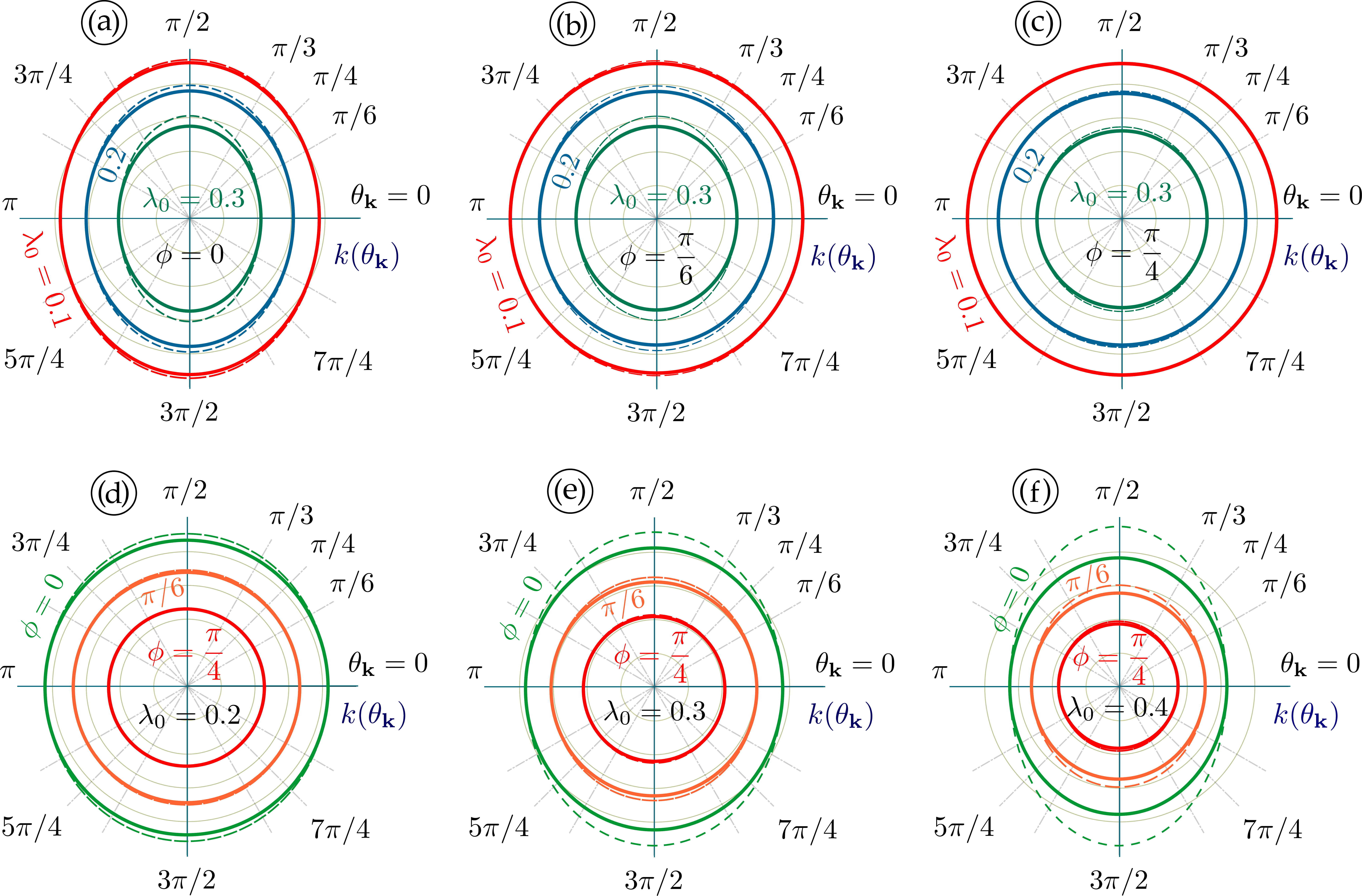}
\caption{(Color online) Polar plots for angular dependence of $\varepsilon_{\alpha}^{\gamma=\pm 1}(\lambda_0,\mbox{\boldmath$k$})$ in Eq.\,\eqref{linD}. 
The upper-row (a)-(c) demonstrates
how $\varepsilon_{\alpha}^{\gamma=\pm 1}(\lambda_0,\mbox{\boldmath$k$})$ depends on the geometry phase $\phi$ with various 
$\lambda_0$ values, where curves for 
$\lambda_0=0.1$ (red), $0.2$ (blue) and $0.3$ (green) are displayed for $\phi=0$ (a), $\pi/6$ (b) and $\pi/4$ (c).
Meanwhile, the lower-row (d)-(f) display $\varepsilon_{\alpha}^{\gamma=\pm 1}(\lambda_0,\mbox{\boldmath$k$})$ dependence on $\lambda_0$ with different 
$\phi$ values, where curves for 
$\phi=0$ (green), $\pi/6$ (orange) and $\pi/4$ (red) are shown for $\lambda_0=0.2$ (d), $0.3$ (e) and $0.4$ (f).
Here, we only present the angular dependence of each dispersion, the size of each oval is not relevant and 
set different for clarity. The dashed (solid) curve corresponds to the exact (expansion, up to the order of ${\cal O}(\lambda_0^2)$) 
calculation of $\mc{F}(\theta_{\bf k} \, \vert \, \phi, \lambda_0)$ in Eq.\,\eqref{TA}.}
\label{FIG:2}
\end{figure}

As displayed in Figs.\,\ref{FIG:2}(d)-\ref{FIG:2}(f), the energy dispersion in Eq.\,\eqref{en} for a dice lattice 
displays an anisotropy due to the applied linearly-polarized light. In general, anisotropic dispersion in Eq.\,\eqref{linD}
can also depend on the phase $\phi$ or parameter $\alpha$ for a general $\alpha-\mc{T}_3$ lattice, as shown in Figs.\,\ref{FIG:2}(a)-\ref{FIG:2}(c). 
This anisotropic effect becomes 
the strongest for graphene but the weakest for a dice lattice by comparing Fig.\,\ref{FIG:2}(a) with Fig.\,\ref{FIG:2}(e) for fixed $\lambda_0=0.3$.  
Numerically, we can verify that Eq.\,\eqref{en}, which is obtained based on the expansion in Eq.\,\eqref{Tmexp}, demonstrates a good accuracy as long as  
$\lambda_0\leq 0.4$. Furthermore, the anisotropy in energy dispersion becomes visible for $\lambda_0 \backsim 0.1$ and above. 
All the ovals are elongated along the $y$ axis for the light polarization along the $x$ direction, and the anisotropy for the dice lattice is manifested as the eccentricity of the 
dispersion ellipses increases with $\lambda_0$ in Figs.\,\ref{FIG:2}(d)-\ref{FIG:2}(f).  
\medskip

In correspondence with the dressed-state energy bands in Eq.\,\eqref{en}, their wave functions are 

\begin{eqnarray}
\label{wfdice1}
&& \Psi^{\gamma = \pm 1}_1 (\tau, \lambda_0, \mbox{\boldmath$k$}) = \frac{1}{k_\lambda} \, 
\left[
\begin{array}{c}
k_x - i \tau k_y \left(1 - \lambda_0^2/4 \right) \\
\sqrt{2}  \, \gamma \, k_\lambda  \\
k_x + i \tau k_y \left(1 - \lambda_0^2/4 \right)
\end{array}
\right] = \frac{1}{k_\lambda} \, 
\left[
\begin{array}{c}
k_x - i \tau a_1(\lambda_0) k_y \\
\sqrt{2}  \, \gamma \, k_\lambda  \\
k_x + i \tau a_1(\lambda_0) k_y
\end{array}
\right] \, , \\
\nonumber 
&& k_\lambda = \frac{1}{\hbar v_F} \, \left|\varepsilon_1^{\gamma=\pm 1} (\lambda_0, \mbox{\boldmath$k$})\right| = \sqrt{k_x^2 + a_1^2(\lambda_0)\,k^2_y}
\end{eqnarray}
Equation\ \eqref{wfdice1} indicates that 
the absolute values of each wave-function component are identical, and therefore, only their phases can be varied by the 
dressing field. This yields

\begin{equation}
\label{wfdice1F}
\Psi_1^{\gamma = \pm 1} (\tau, \lambda_0, \mbox{\boldmath$k$}) = \frac{1}{4} \, 
\left[
\begin{array}{c}
\tet{e}^{- i \Theta^{(1)}_{\bf S} (\tau, {\bf k} \, \vert \, \lambda_0)} \\
\sqrt{2}  \, \gamma  \\
\tet{e}^{+ i \Theta^{(1)}_{\bf S} (\tau, {\bf k} \, \vert \, \lambda_0)}
\end{array}
\right] \, ,
\end{equation}    
where the phase factor is

\begin{equation}
\label{ts0}
\Theta^{(1)}_{\bf S} (\tau, \mbox{\boldmath$k$} \, \vert \, \lambda_0) = \tan^{-1} \left[  
\tau \frac{k_y}{k_x} \, a_1(\lambda_0) \,
\right] =  \tan^{-1} \left[  
\tau a_1(\lambda_0) \, \tan \theta_{\bf k}
\right] 
\end{equation} 
The remaining wave function for the flat band is 
 
\begin{equation}
\label{wfdice0}
\Psi_1^{\gamma=0} (\tau, \lambda_0, \mbox{\boldmath$k$}) = \frac{1}{k_\lambda} \, 
\left[
\begin{array}{c}
k_x - i \tau a_1(\lambda_0) k_y \\
0 \\
-k_x - i \tau a_1(\lambda_0) k_y   
\end{array}
\right] = \left[
\begin{array}{c}
\tet{e}^{- i \Theta^{(1)}_{\bf S}(\tau, {\bf k} \, \vert \, \lambda_0)} \\
0 \\
- \tet{e}^{+ i \Theta^{(1)}_{\bf S} (\tau, {\bf k} \, \vert \, \lambda_0)}
\end{array}
\right] \, .
\end{equation}
Here, all the non-zero wave-function components in Eqs.\,\eqref{wfdice1} and \eqref{wfdice0} have the same absolute value and differ only
by a phase factor in Eq.\,\eqref{ts0}, which is not equal to $\theta_{\bf k}$ if $\lambda_0\neq 0$
and depends on the light intensity, frequency and electron-light coupling.   
\medskip

We also recall the results from Ref.\,[\onlinecite{kisrep}] for graphene ($\alpha = 0$) under linearly-polarized irradiation. The calculated energy 
dispersions are $\varepsilon_0^{\gamma = 0}(\lambda_0, {\bf k}) \equiv 0$ and 

\begin{eqnarray}
\label{DisG}
&& \varepsilon_0^{\gamma = \pm 1}(\lambda_0, \mbox{\boldmath$k$}) = \gamma \, \hbar v_F k\, f_\theta(\lambda_0) \ , \\
\nonumber 
&& f_\theta (\lambda_0) = \sqrt{
 \cos^2 \theta_{\bf k} + 
 J^2_0 (2\lambda_0) \, \sin^2\theta_{\bf k}}\ . 
\end{eqnarray}
The anisotropy factor $a_0 (\lambda_0)$ can be found from the relation 
$\varepsilon_0^{\gamma = \pm 1}(\lambda_0, \mbox{\boldmath$k$}) = \gamma \, \hbar v_F\sqrt{k_x^2 + [a_0 (\lambda_0)\,k_y]^2}$ and gives rise to
 
\begin{equation}
\label{a0}
a_0 (\lambda_0) = J_0(2 \lambda_0) \approx 1- \lambda_0^2 + \frac{\lambda_0^4}{4} + \cdots\ ,
\end{equation}
which matches Eq.\,\eqref{aphi} for general $\alpha-\mc{T}_3$ lattice in the graphene limit $\phi\rightarrow 0$, and is expected to 
play a crucial role in the calculation of transmission of dressed electrons in graphene. 
\medskip

The wave functions associated with Eq.\,\eqref{DisG} for valence and conduction electrons are

\begin{eqnarray}
\label{WFg}
&& \Psi_0^{\gamma = \pm 1}(\lambda_0,\mbox{\boldmath$k$}) = \frac{1}{\sqrt{2}} \, \left[
\begin{array}{c}
 1 \\
 \gamma \, \tet{exp} \left[ i \, \Theta^{(0)}_{\bf S}(\mbox{\boldmath$k$}\,\vert\,\lambda_0) \right]
\end{array}
\right] \ , \\
\nonumber 
&& \Theta^{(0)}_{\bf S}(\mbox{\boldmath$k$}\,\vert\,\lambda_0) = 2 \tan^{-1} \left[ \frac{
\sin \theta_{\bf k} \, J_0(2\lambda_0)
}{
\cos \theta_{\bf k} + f_\theta
} \right] \backsimeq  \theta_{\bf k} - \frac{\lambda_0^2}{2} \, \sin (2 \theta_{\bf k}) + \cdots \ . 
\end{eqnarray}
The wave functions in Eq.\,\eqref{WFg} are somewhat similar to those in Eq.\,\eqref{wfdice1F} for a dice lattice. Here, two wave-function components 
in Eq.\,\eqref{WFg} have the same magnitude, but the phase factor is not equal to $\theta_{\bf k}$ and determined as 

\begin{equation}
\label{phase}
\frac{\tan \left[\Theta^{(0)}_{\bf S}(\mbox{\boldmath$k$}\,\vert\,\lambda_0)\right]}{\tan \theta_{\bf k}}\backsimeq 1 - \lambda_0^2 + 2 \lambda_0^4\,
\sin^2 \theta_{\bf k} \neq a_0(\lambda_0) \ ,
\end{equation}
which implies that the simple phase relation $\tan\left[\Theta^{(\alpha)}_{\bf S}(\mbox{\boldmath$k$}\,\vert\,\lambda_0)\right]/\tan \theta_{\bf k}=a_\alpha(\lambda_0)$
becomes a correct description for all photon-dressed $\alpha-\mc{T}_3$ materials under linearly-polarized irradiation up to the order of $\backsim \lambda_0^2$, including two opposite 
limits for graphene and dice lattice. 
An important difference of graphene, however, is that its wave functions have no explicit dependence on the valley index $\tau = \pm 1$.   
\medskip

Generally speaking, the boundary conditions for dice lattice with $V(x)\neq 0$, as discussed in the next Section \ref{sec3}, 
depend substantially on $k_x$ terms which become discontinuous at two edge boundaries of a potential-barrier region. If the polarization direction of incident light 
lies away from the $y$ direction, such a discontinuity appears in the dressed-state Hamiltonian, as demonstrated in Appendix\ \ref{apb}.
Therefore, {\it the wave-function boundary condition for a dice lattice must be modified accordingly if the energy dispersions become anisotropic}.

\section{Modified Electron Tunneling in Irradiated Graphene and Dice Lattices}
\label{sec3}

The calculated energy dispersions of graphene and dice lattices with $V(x)=0$ in Sec.\,\ref{sec2} under linearly-polarized irradiation,
as well as their wave functions, can now be employed to study the electron transmission dynamics through a square potential barrier. Our main focus 
stays on how the anomalous Klein paradox, i.e., an asymmetrical complete electron transmission which does not depend on the barrier height 
or width, is modified by anisotropic energy dispersion resulted from the applied dressing field. 

\subsection{Anisotropic Dressed-State Tunneling in Graphene}
\label{sec3-1}

We first consider electron tunneling in {\it irradiated graphene} with anisotropic dispersion in Eq.\,\eqref{DisG} as well as a 
two-component wave function in Eq.\,\eqref{WFg}. To some extent, our model system is similar to the asymmetrical tunneling in multi-layer phosphorene with anisotropic dispersion\,\cite{LiuM}, 
where a complete Klein transmission was found at a finite incident angle and termed as the anomalous Klein paradox. 
Physically, however, our system possesses some unique distinctions since the anisotropy factor of irradiated graphene relies on the intensity 
of imposed light (i.e.,, the electron-light coupling $\lambda_0$), and therefore, could be tuned within the off-resonance 
limit $0 < \lambda_0 < 1$. In addition, the polarization direction of incident light could also be varied, instead of being parallel to that
of a head-on electron collision, which is similar to a rotation of the phosphorene larger crystal axis away from the normal direction of a potential barrier. 

\medskip

\begin{figure} 
\centering
\includegraphics[width=0.9\textwidth]{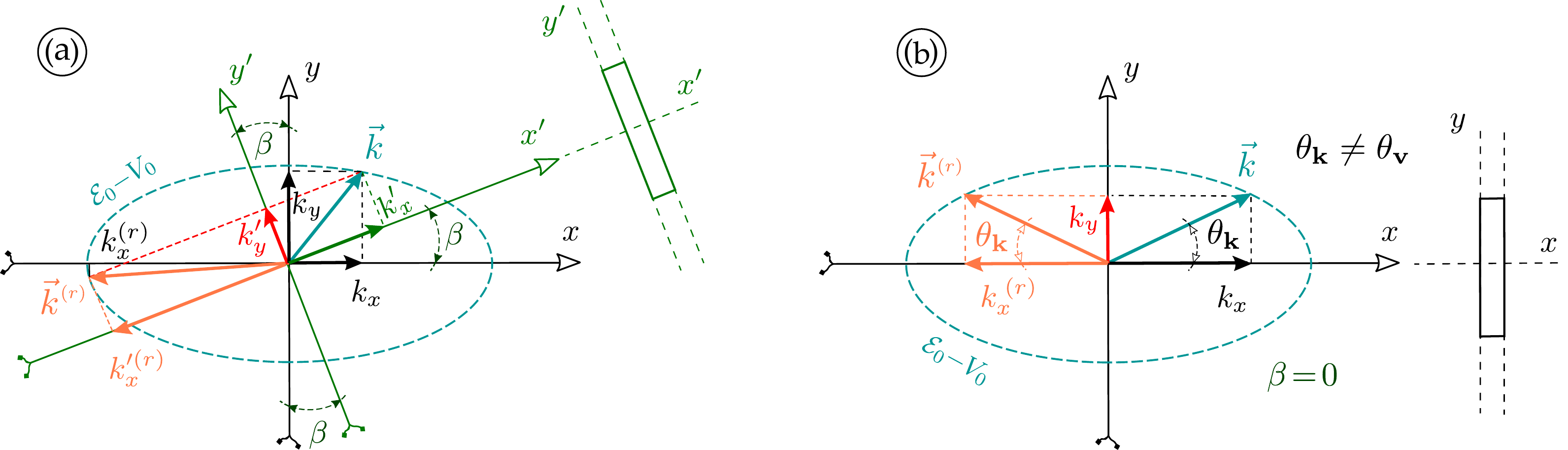}
\caption{(Color online) Schematics for the components of angular momentum {\bf k} and anisotropic energy dispersions of $\alpha-\mc{T}_3$ lattices under linearly-polarized irradiation. Two frames $\{x,y\}$ and $\{x',y'\}$ are associated with the long-axis $\hat{\mbox{\boldmath$x$}}$ of elliptical energy dispersion and the normal-direction $\hat{\mbox{\boldmath$x$}}'$ of potential barrier, respectively.	These two frames are connected to each other by an in-plane rotation angle $\beta$. Here, $\mbox{\boldmath$V$}_G$ and $\mbox{\boldmath$k$}$ are generally not aligned ($\theta_{\bf k}\neq\theta_{\bf V}$) and the panels (a), (b) correspond to $\beta\neq 0$ and $\beta=0$.}
\label{FIG:3}
\end{figure}

In the presence of anisotropic energy dispersion, we find the directions of the group velocity $\mbox{\boldmath$V$}_G$
and the spinor vector $\mbox{\boldmath$S$}$ are aligned neither with each other nor with the electron wave vector 
$\mbox{\boldmath$k$}$, and they are given by

\begin{eqnarray}
\label{SV}
&& \mbox{\boldmath$S$}^{\gamma}(\lambda_0, \mbox{\boldmath$k$}) = \frac{\gamma}{\sqrt{k_x^2 + [a_0(\lambda_0) k_y]^2}} 
\left[
\begin{array}{c} 
k_x \\
a_0(\lambda_0) \, k_y
\end{array} \right]
\ , \\
\label{V}
&& \mbox{\boldmath$V$}_G^{\gamma}(\lambda_0, \mbox{\boldmath$k$}) =  \frac{1}{\hbar} \, \left[ \begin{array}{c} 
\pr/\pr k_x \\
\pr/\pr k_y
\end{array} \right] 
\varepsilon^{\gamma}_0(\lambda_0, \mbox{\boldmath$k$})
= \frac{\gamma\,v_F}{\sqrt{k_x^2 + [a_0(\lambda_0) k_y]^2}}
\left[
\begin{array}{c} 
k_x \\
a^2_0(\lambda_0)\, k_y
\end{array} 
\right] \ .
\end{eqnarray}
Here, the vector $\mbox{\boldmath$S$}^{\gamma}$ is proportional to the spinor wave function in Eq.\,\eqref{WFg}, which switches its direction
to the opposite for the hole state ($\gamma = -1$) compared with the electron state ($\gamma = +1$), while $\mbox{\boldmath$V$}_G^{\gamma}$ characterizes the direction 
of incident particles. 
The corresponding angles of two vectors in Eqs.\,\eqref{SV} and \eqref{V} relative to the $x$-axis are determined as 

\begin{eqnarray}
\label{SVa}
&& \tan \Theta_{\bf S}(\lambda_0) = \left(\frac{k_y}{k_x}\right)\,a_0(\lambda_0)\ , \\
\nonumber
&& \tan \Theta_{\bf V}(\lambda_0) = \left(\frac{k_y}{k_x}\right)\,a_0^2(\lambda_0)\ ,
\end{eqnarray}
or alternatively,
$\tan \Theta_{\bf V}(\lambda_0) = a_0(\lambda_0) \, \tan \Theta_{\bf S}(\lambda_0) = a_0^2(\lambda_0) \, \tan \theta_{\bf k}$.
Since the long-axis of energy dispersion (or $\hat{\mbox{\boldmath$x$}}$ direction) and the normal direction of potential-barrier (or $\hat{\mbox{\boldmath$x$}}'$ direction) 
are generally not aligned to each other, we introduce two coordinate frames:
$\{x,y\}$ for the $\hat{\mbox{\boldmath$x$}}$ vector and $\{x',y'\}$ for the $\hat{\mbox{\boldmath$x$}}'$ vector, as depicted in Fig.\,\ref{FIG:3}. 
Therefore, the same incident-electron wave vector $\mbox{\boldmath$k$}$ can be decomposed either as $\{k_x, k_y\}$ or as $\{k_{x'}, k_{y'}\}$ 
in two different frames but with the same magnitude $k$.
These two frames are related to each other by an in-plane rotation angle $\beta$, as shown in Fig.\,\ref{FIG:3}, and the 
rotation matrix $\hat{\mbb{R}}(\beta)$ is 

\begin{equation}
\hat{\mbb{R}}(\beta) = \left[
\begin{array}{cc}
\cos \beta & -\sin \beta \\
\sin \beta & \cos \beta
\end{array}
\right] \ . 
\label{rota}
\end{equation}
As a result, we have $\theta_{\bf k}=\theta_{{\bf k}'}+\beta$ or $\Theta_{\bf V} = \Theta_{{\bf V}'} + \beta$ and the wave vector vector $\mbox{\boldmath$k$}$ in two frames 
are connected by 

\begin{equation}
\label{BetaRot}
\left[ \begin{array}{c}
k_{x} \\
k_{y}
\end{array}  \right] = 
\left[
\begin{array}{cc}
\cos \beta & -\sin \beta \\
\sin \beta & \cos \beta
\end{array}
\right] 
\left[
\begin{array}{c}
k_{x'} \\
k_{y'}
\end{array} \right]\ .
\end{equation}
Relation in Eq.\,\eqref{BetaRot} holds true for the incident and reflected waves in both barrier and zero-potential regions. 
The reason for introducing another frame $\{x',y'\}$ is the conservation of the transverse wave number $k_{y'}$ across the barrier for all regions. 
\medskip

All wave functions, including the incoming, reflected (with amplitudes $b$ and $r$) and transmitted (with amplitudes $a$ and $t$) waves, could be written explicitly out
in three individual regions as schematically shown in Fig.\,\ref{FIG:1}. For region-$(1)$, we have

\begin{equation}
\label{Psi1G}
\Psi^{(1)}_\gamma (\lambda_0, \mbox{\boldmath$k$}) = \frac{1}{\sqrt{2}}\,
\tet{exp}\left( i k_{x'}^{(1)}x' \right) \tet{exp}\left( i k_{y'}y' \right)\, 
\left[
\begin{array}{c}
1 \\
\gamma\,\tet{e}^{i \Theta^{(1)}_{\bf S}}
\end{array}
\right]  
+ \frac{r}{\sqrt{2}}\, \tet{exp}\left( i k_{x'}^{(1,r)}x' \right) \tet{exp}\left( i k_{y'}y' \right) \, 
\left[
\begin{array}{c}
1 \\
\gamma\,\tet{e}^{i \Theta^{(1,r)}_{\bf S}}
\end{array}
\right]   \ .
\end{equation}
For region-$(2)$ inside the barrier we get

\begin{equation}
\label{Psi2G}
\Psi^{(2)}_{\gamma'} (\lambda_0,  \mbox{\boldmath$k$}) = \frac{a}{\sqrt{2}}\,\tet{exp}\left( i k_{x'}^{(2)}x' \right) \tet{exp}\left( i k_{y'}y' \right)\,
 \left[
\begin{array}{c}
1 \\
\gamma'\,\tet{e}^{i \Theta^{(2)}_{\bf S}}
\end{array}
\right]  + 
\frac{b}{\sqrt{2}} \,\tet{exp}\left( i k_{x'}^{(2,r)}x' \right) \tet{exp}\left( i k_{y'}y' \right)
 \left[
\begin{array}{c}
1 \\
\gamma'\,\tet{e}^{i \Theta^{(2,r)}_{\bf S}}
\end{array}
\right]  \ .
\end{equation}
Finally, for region-$(3)$ we find

\begin{equation}
\label{Psi3G}
\Psi^{(3)}_\gamma (\lambda_0, \mbox{\boldmath$k$}) = \frac{t}{\sqrt{2}}\, \tet{exp}\left( i k_{x'}^{(1)} x' \right) \tet{exp}\left( i k_{y'}y' \right)
\left[
\begin{array}{c}
1 \\
\gamma \, \tet{e}^{i \Theta^{(1)}_{\bf S}}
\end{array}
\right]  \ .
\end{equation}
In Eqs.\,\eqref{Psi1G}-\eqref{Psi3G}, we use the superscripts ``${(1)}$'' and ``${(2)}$'' to denote the wave numbers and angles
in the regions with $V_B=0$ and $V_B>0$, respectively. Meanwhile, the superscript ``${(r)}$'' is used for the backward reflected 
wave in contrast to the incoming and transmitted forward-going waves. Here, all the wave vector components $k_{x'}$ and $k_{y'}$ are given in the 
$\{x',y'\}$ frame, while the spinor angle $\Theta_{\bf S}$ and the group-velocity angle $\Theta_{\bf V}$ must be defined in the $\{x, y\}$ frame related to long-axis of elliptical energy dispersion.
\medskip

The four unknown amplitudes $a$, $b$, $r$ and $t$ in Eqs.\,\eqref{Psi1G}-\eqref{Psi3G} can be determined from the proper boundary conditions at $x'=0$ and $x'=W_B$. 
We have derived such conditions in Appendix \ref{apd} by integrating all the components of the eigenvalue equation for the corresponding Hamiltonian over a small interval
$[-\delta x', \delta x']$ and taking the limit of $\delta x' \rightarrow 0$ afterwards. 
\medskip

For graphene with an anisotropic dispersions in Eq.\,\eqref{DisG}, the obtained boundary conditions are found the same as those for isotropic case, and therefore,
we match the two components of the wave-function spinor at both edges of region-$(2)$.\,\cite{LiuM} 
This leads to the following four equations

\begin{eqnarray}
\nonumber
&& 1 + r = a + b \, , \\
\nonumber 
&& \tet{e}^{i \, \Theta_{\bf S}^{(1)}} + r \, \tet{e}^{i \, \Theta_{\bf S}^{(1,r)}} = 
s\left(a \, \tet{e}^{i \, \Theta_{\bf S}^{(2)}} + b \, \tet{e}^{i \, \Theta_{\bf S}^{(2,r)}}\right)\ , \\
\nonumber 
&& a  \, \tet{e}^{i \, k_{x'}^{(2)}W_B} + b \, \tet{e}^{i \, k_{x'}^{(2,r)}W_B} = 
s \, t \, \tet{e}^{i \, k_{x'}^{(1)} W_B} \ , \\
\label{MainSysTG}
&& a  \, \tet{e}^{i \, \Theta_{\bf S}^{(2)} + i \, k_{x'}^{(2)}W_B} + b \, \tet{e}^{i \, \Theta_{\bf S}^{(2,r)} + i \, k_{x'}^{(2,r)}W_B}= 
s \, t \, \tet{e}^{i \, \Theta_{\bf S}^{(1)} + i \, k_{x'}^{(1)}W_B} \ ,
\end{eqnarray}
where $s = \gamma\,\gamma' = \text{sgn}(\varepsilon_0)\,\text{sgn}(\varepsilon_0 - V_B)  = \pm 1$ is the composite index characterizing the same or different
electron-hole transitions at two boundary edges, and $\text{sgn}(x)=+1$ or $-1$ for $x>0$ or $x<0$. Furthermore,
the transmission and reflection coefficients are obtained as $T(\varepsilon_0,\Theta_{{V}_{x'}}^{(1)} \, \vert \, \beta)=\vert t \vert ^2$ 
and $R(\varepsilon_0,\Theta_{{V}_{x'}}^{(1)} \, \vert \, \beta)=\vert r \vert ^2$, respectively, satisfying the relation 
$R(\varepsilon_0,\Theta_{{V}_{x'}}^{(1)} \, \vert \, \beta) = 1 - T(\varepsilon_0,\Theta_{{V}_{x'}}^{(1)} \, \vert \, \beta)$. 
\medskip

In order to solve four boundary equations in Eq.\,\eqref{MainSysTG}, we need find the spinor angles for the incoming and reflected waves, 
both outside $\Theta_{\bf S}^{(1)}$, $\Theta_{\bf S}^{(1,r)}$ and inside $\Theta_{\bf S}^{(2)}$, $\Theta_{\bf S}^{(2,r)}$ the barrier region. 
These spinor angles are decided by $k_x$ and $k_y$ in the $\{x,y\}$ frame, 
while the electron wave numbers $k_{x'}^{(r)}$, $k_{x'}^{(1,r)}$, $k_{x'}^{(2)}$, $k_{x'}^{(2,r)}$ and $k_{y'} = \text{const}$ in Eq.\,\eqref{MainSysTG} are given in the $\{x',y'\}$ frame. 
For the whole tunneling process, the given parameters are the kinetic energy of the incoming particle $\varepsilon_0$ as well as 
the angle $\Theta_{{V}_{x'}}^{(1)}$ between its group velocity vector and the $x'$ axis, i.e., the direction of the incoming particles with respect to the normal direction of barrier barrier. 
\medskip

We first notice that all the unknowns involved in Eq.\,\eqref{MainSysTG} are associated with both $\{x,y\}$ and $\{x',y'\}$ frames.
Using the given kinetic energy $\varepsilon_0$ for incident particles in region-($1$), from Eq.\,\eqref{DisG} we first find $\{k_{x}^{(1)}, k_{y}^{(1)}\}$ in the $\{x,y\}$ frame, i.e.,

\begin{equation}
\label{Ek}
\left[k_{x}^{(1)}\right]^2 + \left[ a_0(\lambda_0) \, k_{y}^{(1)} \right]^2 = \left( \frac{\varepsilon_0}{\hbar v_F} \right)^2 \ .
\end{equation}
Similarly, in region-($2$) we get 

\begin{eqnarray}
\label{Ek2}
&& \left[k_{x}^{(2)}\right]^2 + \left[ a_0(\lambda_0) \, k_{y}^{(2)} \right]^2 = \left(\frac{\varepsilon_0 - V_B}{\hbar v_F} \right)^2 \ .
\end{eqnarray}
Knowing $\{k_x^{(1)}, k_y^{(1)}\}$ and $\{k_x^{(2)}, k_y^{(2)}\}$ in the $\{x,y\}$ frame, we are able to find $\Theta_{\bf S}^{(1)}$ and $\Theta_{\bf S}^{(2)}$ easily from 
Eq.\,\eqref{SVa}.
\medskip

Physically, it is the group-velocity component $V_{G, x'}^\gamma$ in Eq.\,\eqref{V}, or its angle $\Theta_{{\bf V}'}$ in Eq.\,\eqref{SVa}, within the $\{x',y'\}$ frame that determines 
the direction of a moving wave.\,\cite{LiuM} 
We know that there exist two solutions within the $\{x',y'\}$ frame from either Eq.\,\eqref{Ek} or Eq.\,\eqref{Ek2} in region-($1$) and region-($2$), respectively,
corresponding to the forward ($V_{G, x'}^\gamma>0$) and backward ($V_{G, x'}^\gamma<0$) moving waves, respectively. The frame-rotation matrix in Eq.\,\eqref{rota} 
can project these found solutions $\{k^{(1,2)}_{x'}, k^{(1,2)}_{y'}\}_\pm$ back to $\{k^{(1,2)}_{x,}, k^{(1,2)}_y\}_\pm$ in the $\{x,y\}$ frame 
from which the spinor angles $\Theta_{{\bf S},\pm}^{(1)}$ and $\Theta_{{\bf S},\pm}^{(1,r)}$ in region-($1)$, as well as $\Theta_{{\bf S},\pm}^{(2)}$ and $\Theta_{{\bf S},\pm}^{(2,r)}$ in region-($2)$, 
can be calculated based on Eq.\,\eqref{SVa}.
\medskip

\begin{figure} 
\centering
\includegraphics[width=0.7\textwidth]{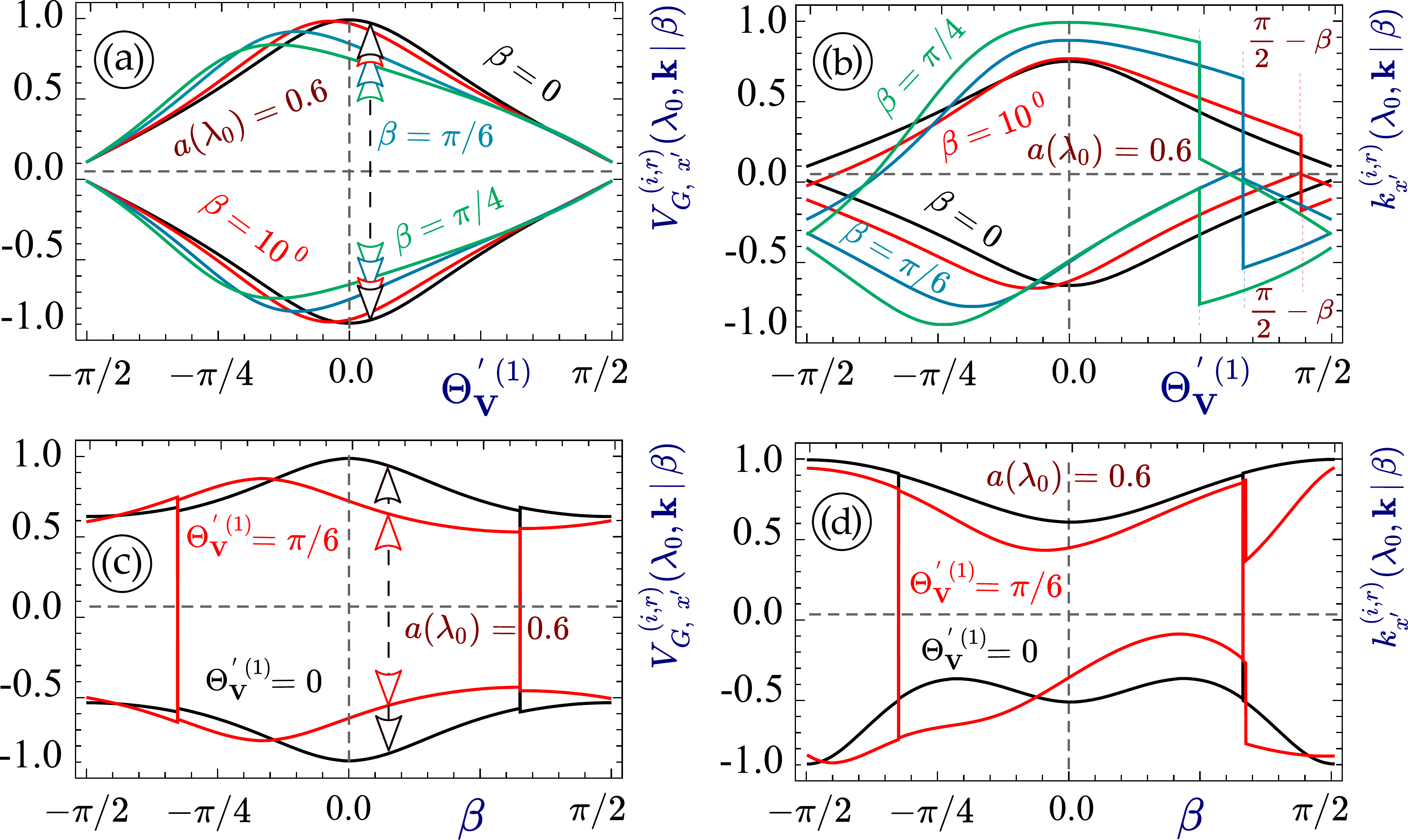}
\caption{(Color online) $V_{G, x'}^{\gamma} (\lambda_0, \mbox{\boldmath$k$}\, \vert \, \beta)$ [$(a)$, $(c)$] from Eq.\,\eqref{Vxp} 
and $k^{(1)}_{x'}$ [$(b)$, $(d)$] solved from Eq.\,\eqref{Ek} for $a_0(\lambda_0)=0.6$ as a function of the incidence angle $\Theta_{{\bf V}'}^{(1)}$ [$(a)$, $(b)$] and 
a function of rotation angle $\beta$ [$(c)$, $(d)$]. 
In all panels, the upper four curves correspond to the incident wave, while the 
lower four are for the reflected one. Two $V_{G, x'}^{\gamma} (\lambda_0, \mbox{\boldmath$k$}\, \vert \, \beta)$ components
for the incident and reflected waves are always with opposite signs, which is, however, not true for two $k^{(1)}_{x'}$ components. 
In panels $(a)$, $(b)$, different curves correspond to $\beta = 0$, $10^{\,\rm o}$, $\pi/6$ and $\pi/4$, while $\Theta_{{\bf V}'}^{(1)} = 0$ and $\pi/6$ are taken
for two curves in plots $(c)$, $(d)$.}
\label{FIG:4}
\end{figure}

Two unknwn components of the group velocity vector $\mbox{\boldmath$V$}_{G}^{\gamma} (\lambda_0, \mbox{\boldmath$k$})$ in the $\{x',y'\}$ frame 
can be obtained from two known components of $\mbox{\boldmath$V$}_{G}^{\gamma} (\lambda_0, \mbox{\boldmath$k$})$ in the $\{x,y\}$ frame
by using Eq.\,\eqref{BetaRot}, yielding

\begin{equation}
\label{Vp}
\left[
\begin{array}{c}
V_{G, x'}^{\gamma} (\lambda_0, \mbox{\boldmath$k$}) \\
V_{G, y'}^{\gamma} (\lambda_0, \mbox{\boldmath$k$})
\end{array}  
\right]=\hat{\mbb{R}}(-\beta)\left[
\begin{array}{c}
V_{G, x}^{\gamma} (\lambda_0, \mbox{\boldmath$k$}) \\
V_{G, y}^{\gamma} (\lambda_0, \mbox{\boldmath$k$})
\end{array}  
\right]
=\hat{\mbb{R}}(-\beta)\left\{\frac{\gamma\,v_F}{\sqrt{k_x^2 + [a_0(\lambda_0)\,k_y]^2}}\,
\left[
\begin{array}{c} 
k_x \\
a^2_0(\lambda_0) \, k_y
\end{array} 
\right]\right\} \ .
\end{equation}
We solve Eq.\,\eqref{Vp} together with Eq.\,\eqref{Ek} for $V_{G, x'}^{\gamma} (\lambda_0, \mbox{\boldmath$k$})$ and express two solutions explicitly through the 
known $k_{y'}$ as       

\begin{equation}
\label{Vxp}
V_{G, x'}^{\gamma} (\lambda_0, \mbox{\boldmath$k$}\, \vert \, \beta)= \pm \gamma \, \frac{v_F}{\sqrt{2}} \, 
\left\{ 
1 + a_0^2(\lambda_0) + \left[ 1 - a_0^2(\lambda_0) \right] \, \cos (2 \beta) \, 
- 2 \left[\frac{\hbar v_F}{\varepsilon_0} \, a_0(\lambda_0) \, k_{y'} \right]^2 \,
\right\}^{1/2} \ ,
\end{equation}
which have the opposite signs and equal magnitudes, indicating one forward ($+$) and one backward ($-$) waves, respectively. 
We emphasize that the relation in Eq.\,\eqref{Vxp} does not hold true for the case of wave
vector components, such as $k_x$ or $k_{x'}$, as demonstrated in Fig.\,\ref{FIG:4}. In the absence of rotation ($\beta = 0$) between two frames, Eq.\,\eqref{Vxp} reduces to 

\begin{equation}
V_{G, x'}^{\gamma} (\lambda_0, \mbox{\boldmath$k$}\, \vert \, \beta\to 0) = \pm \gamma \, v_F \, 
\sqrt{1 -\left[ \frac{\hbar v_F}{\varepsilon_0} \, a_0(\lambda_0) \, k_{y'} \right]^2 } \ ,
\label{rela1}
\end{equation}
or 

\begin{equation}
V_{G, x'}^{\gamma} (\lambda_0\to 0, \mbox{\boldmath$k$}\, \vert \, \beta) = \pm \gamma \, v_F \, 
\sqrt{1 -\left(\frac{\hbar v_F}{\varepsilon_0} \, k_{y'} \right)^2 }
\label{rela2}
\end{equation}
if the electron-light interaction and anisotropy are turned off with $\lambda_0=0$, which is independent of angle $\beta$.
\medskip

The perfect transmission is achieved\,\cite{LiuM} if $\mbox{\boldmath$k$} =\{ k \cos \beta,\, k \sin \beta \}$  or $\beta = \theta_{\bf k}$ 
($\mbox{\boldmath$k$}\|\hat{\mbox{\boldmath$x$}}'$), which implies
$\theta'_{\bf k} = \theta_{\bf k} - \beta = 0$. In this case, however, 
the angle of incidence $\Theta_{{\bf V}'}$ for the 
perfect transmission is not equal to $- \beta$ as expected, but is found to be

\begin{equation}
\label{vpt}
\Theta_{{\bf V}'} = \Theta_{\bf V} - \beta = \tan^{-1} \left[ a_0^2(\lambda_0)\tan \theta_{\bf k} \right] - \beta
= \tan^{-1} \left[ a_0^2(\lambda_0) \tan \beta \right] - \beta\ .
\end{equation}
Therefore, the maximum possible value\,\cite{LiuM} of $\Theta^{\rm max}_{{\bf V}'}$ in Eq.\,\eqref{vpt} can be reached when $\beta=\beta_0=\tan^{-1}[1/a_0(\lambda_0)]$, leading to 

\begin{equation}
\label{ThMax}
\Theta^{\rm max}_{{\bf V}'} =\tan^{-1} \left( 1- \lambda_0^2 \right)  -
 \tan^{-1} \left( \frac{1}{1- \lambda_0^2} \right) \backsimeq - \lambda_0^2 - \frac{\lambda_0^4}{2} - \frac{\lambda_0^6}{6} + \cdots\ ,
\end{equation}
which holds true for both graphene and a dice lattice. 

\par 
\medskip
Even though we do not consider any collective effects or Fermi surfaces here, it is convenient to express all our quantities and their
numerical values in terms of a single unit corresponding to a typical Fermi momentum of a graphene electron. Such momentum is equal to $k_F^{\,(0)} \backsim 10^6\,cm^{-1}$ for a standard two-dimensional electron density $n_0 = 1.0 \cdot 10^{11}\,cm^{-2}$. Therefore, we will measure all our lengths, such as barrier widths, in terms of $l_0 = 1/k_F^{\,(0)} \backsimeq 10^{-6}\,cm \backsimeq 10\,nm$, while our unit of energy will be taken as $E_0 = \hbar v_F k_F^{\,(0)} \backsimeq 10 - 100 \,meV$.    

\subsection{Anisotropic Dressed-State Tunneling in Dice Lattice}
\label{sec3-2}

Next, we consider the electron tunneling in an irradiated dice lattice with the anisotropic dispersion in Eq.\,\eqref{linD}. We first note that the geometry of the 
Dirac cones, both isotropic and anisotropic, are exactly the same for graphene and a dice lattice apart from the existence of a flat band. 
Therefore, all the reasoning and derivations in Sec.\,\ref{sec3-1} for electron wave vectors, spinor and group velocity angles {\it are also applicable for a dice lattice}, including 
Eqs.\,\eqref{SV}-\eqref{Ek2}. We will not use the flat-band wave function to avoid the situation with zero kinetic energy in all three regions 
because of its infinite degeneracy of electron wave numbers.       
\medskip

For a dice lattice, we want to find the wave functions in all three regions of Fig.\,\ref{FIG:1}. Specifically, in region-($1$) we have 

\begin{equation}
\label{Psi1D}
\Psi^{(1)}_{1} (\gamma_1,\lambda_0 \, \vert \, \mbox{\boldmath$k$}) = \frac{1}{4} \,
\tet{exp}\left(i k_{x'}^{(1)} \, x' \right) \tet{exp}\left(i k_{y'} \, y'\right) \left[
\begin{array}{c}
\tet{e}^{- i \Theta_{\bf S}^{(1)}}\\
\sqrt{2} \, \gamma_1 \\
\tet{e}^{i \Theta_{\bf S}^{(1)}}
\end{array}
\right] + 
\frac{r}{4} \, \tet{exp}\left(i k_{x'}^{(1,r)} \, x' \right) \tet{exp}\left(i k_{y'} \, y'\right)
\left[
\begin{array}{c}
\tet{e}^{- i \Theta_{\bf S}^{(1,r)}}\\
\sqrt{2} \, \gamma_1 \\
\tet{e}^{i \Theta_{\bf S}^{(1,r)}}
\end{array}
\right] \ ,
\end{equation}
where $\gamma_1=+1$ for electrons and $-1$ for holes and the spinor angle 
$\Theta_{\bf S}(\mbox{\boldmath$k$}\,\vert\,\lambda_0)$ has been presented in Eq.\,\eqref{ts0}. 
Similarly, we find the wave function in region-($2$) as

\begin{equation}
\label{Psi2D}
\Psi^{(2)}_{1} (\gamma'_1, \lambda_0 \, \vert \, \mbox{\boldmath$k$}) = \frac{a}{4} \,
\tet{exp}\left(i k_{x'}^{(2)}x' \right) \tet{exp}\left(i k_{y'}y' \right)
\left[
\begin{array}{c}
\tet{e}^{- i \Theta_{\bf S}^{(2)}}\\
\sqrt{2} \, \gamma'_1 \\
\tet{e}^{i \Theta_{\bf S}^{(2)}}
\end{array}
\right]
+ \frac{b}{4}\,\tet{exp}\left(i k_{x'}^{(2,r)}x'\right) \tet{exp}\left(i k_{y'}y' \right)
\left[
\begin{array}{c}
\tet{e}^{- i \Theta_{\bf S}^{(2,r)}}\\
\sqrt{2} \, \gamma'_1 \\
\tet{e}^{i \Theta_{\bf S}^{(2,r)}}
\end{array}
\right]  \ ,
\end{equation}
and the wave function in region-($3$) to be

\begin{equation}
\label{Psi3D}
\Psi^{(3)}_{\, 1} (\gamma_1, \lambda_0 \, \vert \, \mbox{\boldmath$k$})= \frac{t}{4} \, 
\tet{exp}\left(i k_{x'}^{(1)} x' \right) \tet{exp}\left( i k_{y'} y' \right)
\left[
\begin{array}{l}
\tet{e}^{- i \Theta_{\bf S}^{(1)}}\\
\sqrt{2} \, \gamma_1 \\
\tet{e}^{i \Theta_{\bf S}^{(1)}}
\end{array}
\right] \ .
\end{equation}
Here, it is straightforward to verify that for the case of an isotropic disperion 
with $a_1(\lambda_0) = 1$ and $\beta = 0$, the spinor angle $\Theta_{\bf S}$ is the same as the 
wave-vector angle $\theta_{\bf k}$. Meanwhile, we also acquire $\Theta_{\bf S}^{(1,r)} \rightarrow \pi - \theta_{\bf k}^{(1)}$ and  
$ \Theta_{\bf S}^{(2,r)} \rightarrow \pi - \theta_{\bf k}^{(2)}$.  Furthermore, the electron wave numbers for the forward and backward waves 
become $k_{x'}^{(1,r)} \rightarrow - k_{x}^{(1)}$, $k_{x'}^{(2,r)} \rightarrow - k_{x}^{(2)}$, 
$k_{x'}^{(1)} \rightarrow k_{x}^{(1)}$ and $k_{y'} \rightarrow k_{y}$. Finally, Eq.\,\eqref{Ek} gives rise to
$k_{x}^{(1)} \rightarrow \sqrt{\left(\varepsilon_0/\hbar v_F\right)^2 - k_y^2}$. In this way, all the obtained 
expressions in Sec.\,\ref{sec3-1} for the electron tunneling in graphene\,\cite{neto, kacm} can be transformed into corresponding ones for a dice lattice\,\cite{alphaDice}.   
\medskip

The composite boundary conditions for an anisotropic dice lattice, as derived in Appendix\ \ref{apd}, are different from 
simple boundary conditions for graphene (i.e., we cannot simply match individual component of the wave function) or even not equivalent to the boundary conditions for
an isotropic dice lattice since the additional $k_{x'}$-dependent term in the dressed-state Hamiltonian leads to 
additional instances of discontinuity at the boundaries of the barrier region. The new composite boundary conditions are found to be  

\begin{equation}
\varphi_2(-\delta x') = \varphi_2(\delta x')\ ,
\end{equation}
and 

\begin{equation}
c_\tau^+(\lambda_0, \beta) \, \varphi_1(- \delta x') +  c_\tau^-(\lambda_0, \beta) \, \varphi_3(- \delta x') = 
c_\tau^+(\lambda_0, \beta) \, \varphi_1(\delta x') + c_\tau^-(\lambda_0, \beta) \, \varphi_3(\delta x') \ ,
\end{equation}
where index $j=1,\,2,\,3$ labels three components of wave functions, and

\begin{equation}
c_\tau^\pm (\lambda_0, \beta) = \tau \cos \beta \pm i \, a_1(\lambda_0) \sin \beta \ .
\end{equation} 
Here, {\it the modified boundary conditions for an anisotropic dice lattice represent one of the key results of the present paper}. 
Consequently, the boundary equations
to determined the transmission and reflections amplitudes are given explicitly by

\begin{eqnarray}
\label{MainSystD2}
&& 1 + r = s (a + b) \, , \\
\nonumber 
&&  \cos \left[ \Theta_{\bf S}^{(1)} - \alpha_\tau (\beta, a_1) \right] + 
r \, \cos \left[ \Theta_{\bf S}^{(r,1)} - \alpha_\tau (\beta, a_1) \right] = 
a\, \cos \left[ \Theta_{\bf S}^{(2)} - \alpha_\tau (\beta, a_1) \right] + 
b\, \cos \left[ \Theta_{\bf S}^{(r,2)} - \alpha_\tau (\beta, a_1) \right] \ , \\
\nonumber 
&& a \, \tet{exp}\left(i \, k_{x'}^{(2)} \, W_B \right) + 
b \, \tet{exp}\left(i \, k_{x'}^{(2,r)} \, W_B \right) = 
t \, s \, \tet{exp}\left(i \, k_{x'}^{(1)} \, W_B \right) \ , \\
\nonumber 
&& a \, \cos \left[ \Theta_{\bf S}^{(2)} - \alpha_\tau (\beta, a_1) \right] \, \tet{exp}\left(i \, k_{x'}^{(2)} \, W_B \right) + 
b \, \cos \left[ \Theta_{\bf S}^{(r,2)} - \alpha_\tau (\beta, a_1) \right] \, \tet{exp}\left(i \, k_{x'}^{(2,r)} \, W_B \right)  
\\ 
\nonumber 
&& = t \, s \, \cos \left[ \Theta_{\bf S}^{(1)} - \alpha_\tau (\beta, a_1) \right] \, \tet{exp}\left(i \, k_{x'}^{(1)} \, W_B \right) \ ,
\end{eqnarray}
where $\alpha_\tau (\beta, a_1) = \tau \, \tan^{-1} \left[ a_1(\lambda_0) \, \tan \beta \right]$, $s = \gamma_1 \gamma'_1 = \text{sign}(\varepsilon_0)\,\text{sign}(\varepsilon_0 - V_B)  = \pm 1$ is 
a composite index characterizing a possible electron $\rightarrow$ hole $\rightarrow$ electron transition, 
similarly to what we have obtained for graphene. Once $\mbox{\boldmath$k$}$ and $\Theta_{\bf S}$ are known, the transmission and reflection amplitudes, 
$t$ and $r$, could be calculated in a straightforward way. 

\section{Results and Discussions} 
\label{sec4}

The electron tunneling, transmission and reflection amplitudes are largely determined by the band structure and the property of 
dressed-state wave functions of electrons. For both graphene and dice lattice, the linearly-polarized irradiation modifies the phase factors of individual  
wave-function components. These light-induced modifications are found different for graphene and for dice, 
and especially, they also differ for the valence/conduction and the flat-band wave functions. 
The preserved symmetry leads to the occurrence of anomalous Klein paradox, but not the head-on collisions. 
\medskip

Klein paradox is a complete transmission of the incoming particles independent of the barrier height and width. 
It is quite different from the resonant Fabry-Perot tunneling resulted from a constructive interference with unity peaks on both sides of the Klein maximum.  
The transmission peaks of these two cases could be easily resolved for a standard isotropic Dirac spectrum since the Klein paradox only occurs 
for the head-on collision. In fact, the following approximate expression for the electron transmission\,\cite{neto}

\begin{equation}
\label{tneto}
T_0 \left(\varepsilon_0, \theta_{\bf k}^{(1)} \, \vert \, \beta = 0  \right) \approx \frac{\cos^2 \theta_{\bf k}^{(1)} }{1 - \cos^2
\left( k_x^{(2)} W_B \right) \, \sin^2 \theta_{\bf k}^{(1)}} 
\end{equation}
suggests that a complete transmission is always present for the head-on collision with $\theta_{\bf k}^{(1)} = 0$ or the Klein
paradox, but there also exists a number of other resonances of unimpeded tunneling corresponding to $k_x^{(2)} W_B = \pi \times 
\text{integer}$ with their peak locations depending on the barrier width $W_B$ and the longitudinal wave number $k_x^{(2)}$ within the barrier region.
The latter quantity is determined from the relation involving the kinetic energy $\varepsilon_0$ of incoming particles and the barrier height $V_B$. 
\medskip

\begin{figure} 
\centering
\includegraphics[width=0.9\textwidth]{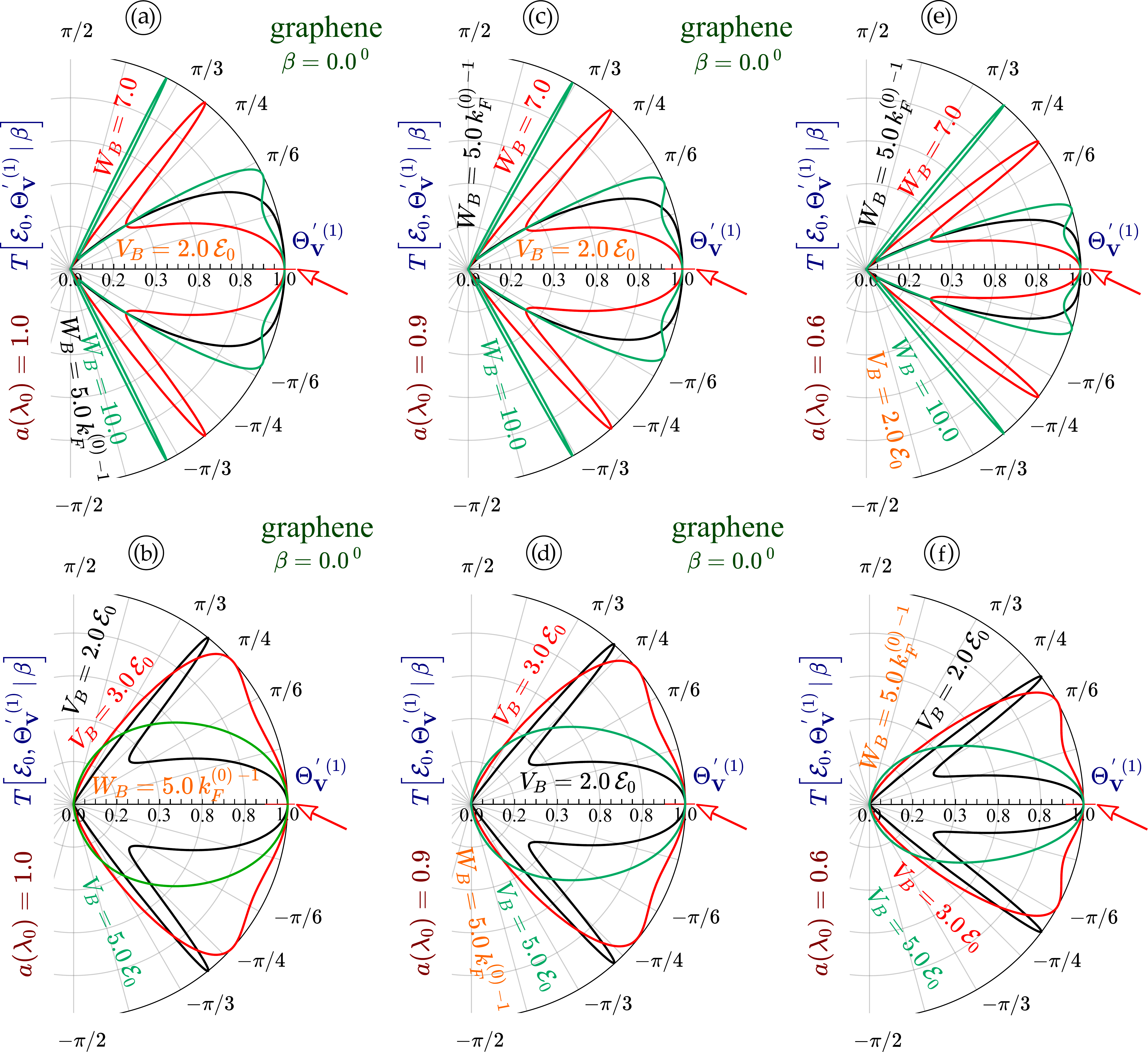}
\caption{(Color online) Angular plots for the electron transmission $T\left(\varepsilon_0, \Theta_{{\bf V}'}^{(1)}\, \vert \,\beta \right)$ 
as a function of the incident angle $\Theta_{{\bf V}'}^{(1)}$ in Eq.\,\eqref{Vp} for the group-velocity direction of incident electrons in graphene. 
Each panel relates to a specific value of $a_0(\lambda_0)$ in Eq.\,\eqref{DisG}: $1.0$ for $(a)$, $(b)$; $0.9$ for $(c)$, $(d)$; $0.6$ for $(e)$, $(f)$.
Panels $(a)$, $(c)$, $(e)$ demonstrate the transmission by black, red and green curves for $k_F^{(0)}W_B = 5,\,7,\,10$ and $V_B/\varepsilon_0=2$, as well as for 
$V_B/\varepsilon_0 = 2,\,3,\,5$ and $k_F^{(0)}W_B=5$ in $(b)$, $(d)$, $(f)$.
The direction of the shifted non-head-on Klein paradox is specified by the same complete transmission disregarding of the barrier height or width and indicated by the red arrow in each panel. 
Here, $\beta=0^{\,\rm o}$ is set in all panels for the angle between $k_{x'}$ of the surface normal of barrier and $k_x$ for the long-axis of elliptical energy dispersion.}
\label{FIG:5}
\end{figure}

\begin{figure} 
\centering
\includegraphics[width=0.9\textwidth]{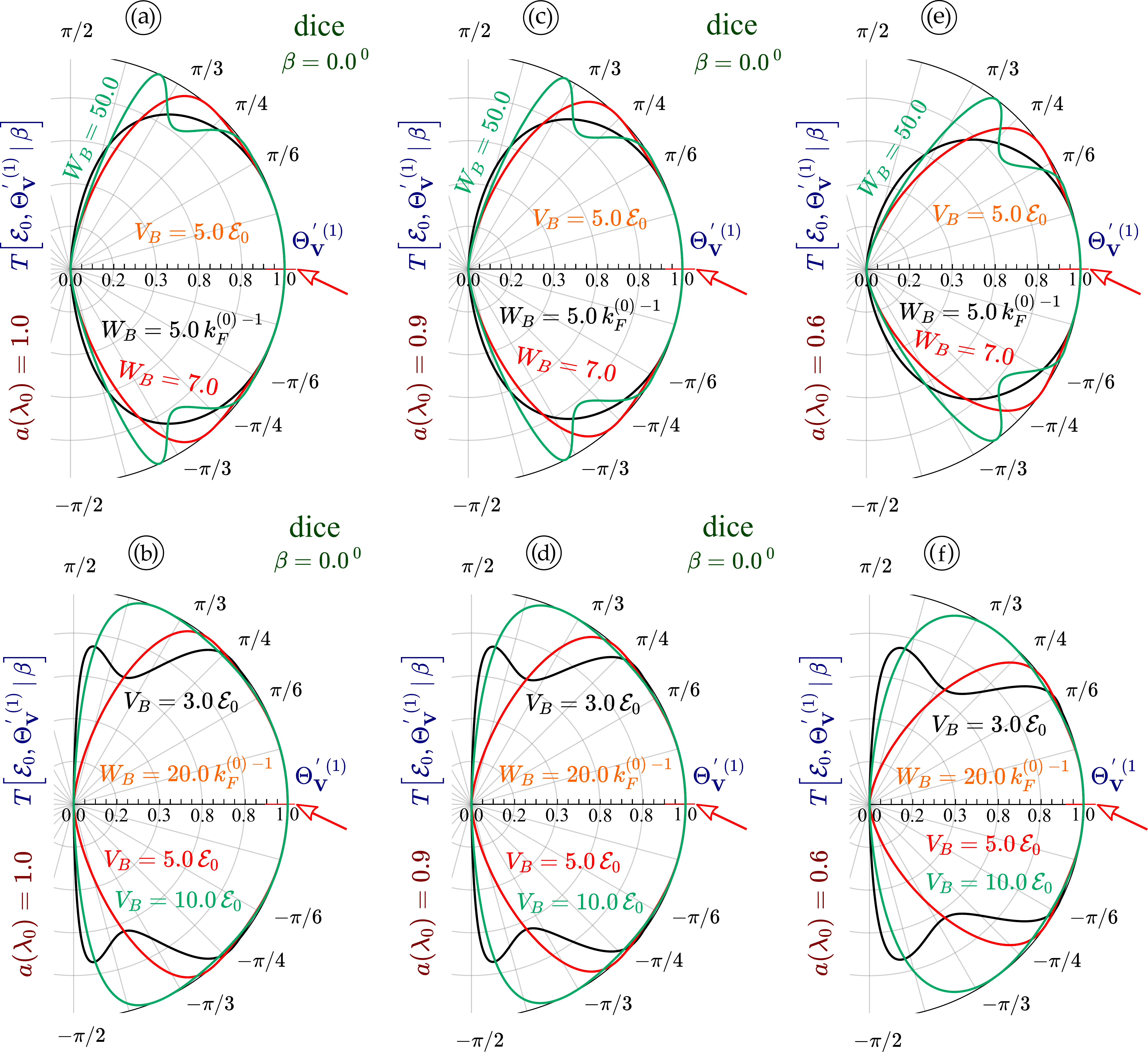}
\caption{(Color online) Angular plots for $T\left(\varepsilon_0, \Theta_{{\bf V}'}^{(1)}\, \vert \,\beta \right)$ 
as a function of $\Theta_{{\bf V}'}^{(1)}$ in dice lattices. 
Each panel relates to a specific value of $a_1(\lambda_0)=1.0$ for $(a)$, $(b)$; $0.9$ for $(c)$, $(d)$; $0.6$ for $(e)$, $(f)$.
Panels $(a)$, $(c)$, $(e)$ demonstrate the transmission by black, red and green curves for $k_F^{(0)}W_B = 5,\,7,\,50$ and $V_B/\varepsilon_0=5$, as well as for 
$V_B/\varepsilon_0 = 3,\,5,\,10$ and $k_F^{(0)}W_B=20$ in $(b)$, $(d)$, $(f)$.
The direction of the shifted non-head-on Klein paradox is indicated by the red arrow in each panel. 
Here, $\beta=0^{\,\rm o}$ is set for all panels.}	
\label{FIG:11}
\end{figure}

\begin{figure} 
\centering
\includegraphics[width=0.9\textwidth]{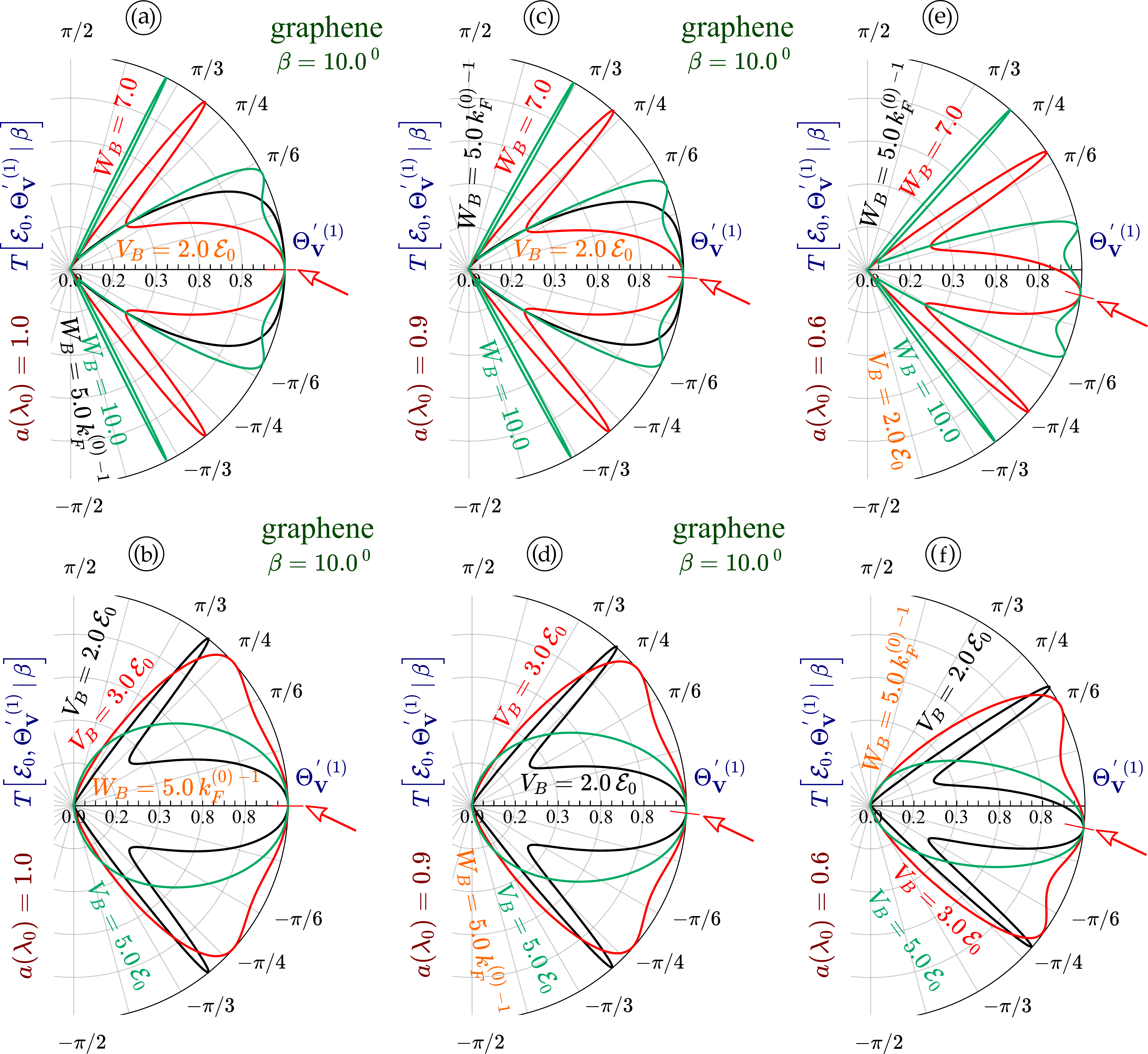}
\caption{(Color online) Angular plots for $T\left(\varepsilon_0, \Theta_{{\bf V}'}^{(1)}\, \vert \,\beta \right)$ 
as a function of $\Theta_{{\bf V}'}^{(1)}$ in graphene. 
Each panel relates to a specific value of $a_0(\lambda_0)=1.0$ for $(a)$, $(b)$; $0.9$ for $(c)$, $(d)$; $0.6$ for $(e)$, $(f)$.
Panels $(a)$, $(c)$, $(e)$ demonstrate the transmission by black, red and green curves for $k_F^{(0)}W_B = 5,\,7,\,10$ and $V_B/\varepsilon_0=2$, as well as for 
$V_B/\varepsilon_0 = 2,\,3,\,5$ and $k_F^{(0)}W_B=5$ in $(b)$, $(d)$, $(f)$.
The direction of the shifted non-head-on Klein paradox is indicated by the red arrow in each panel. 
Here, $\beta=10^{\,\rm o}$ is set for all panels.}
\label{FIG:6}
\end{figure}

\begin{figure} 
\centering
\includegraphics[width=0.9\textwidth]{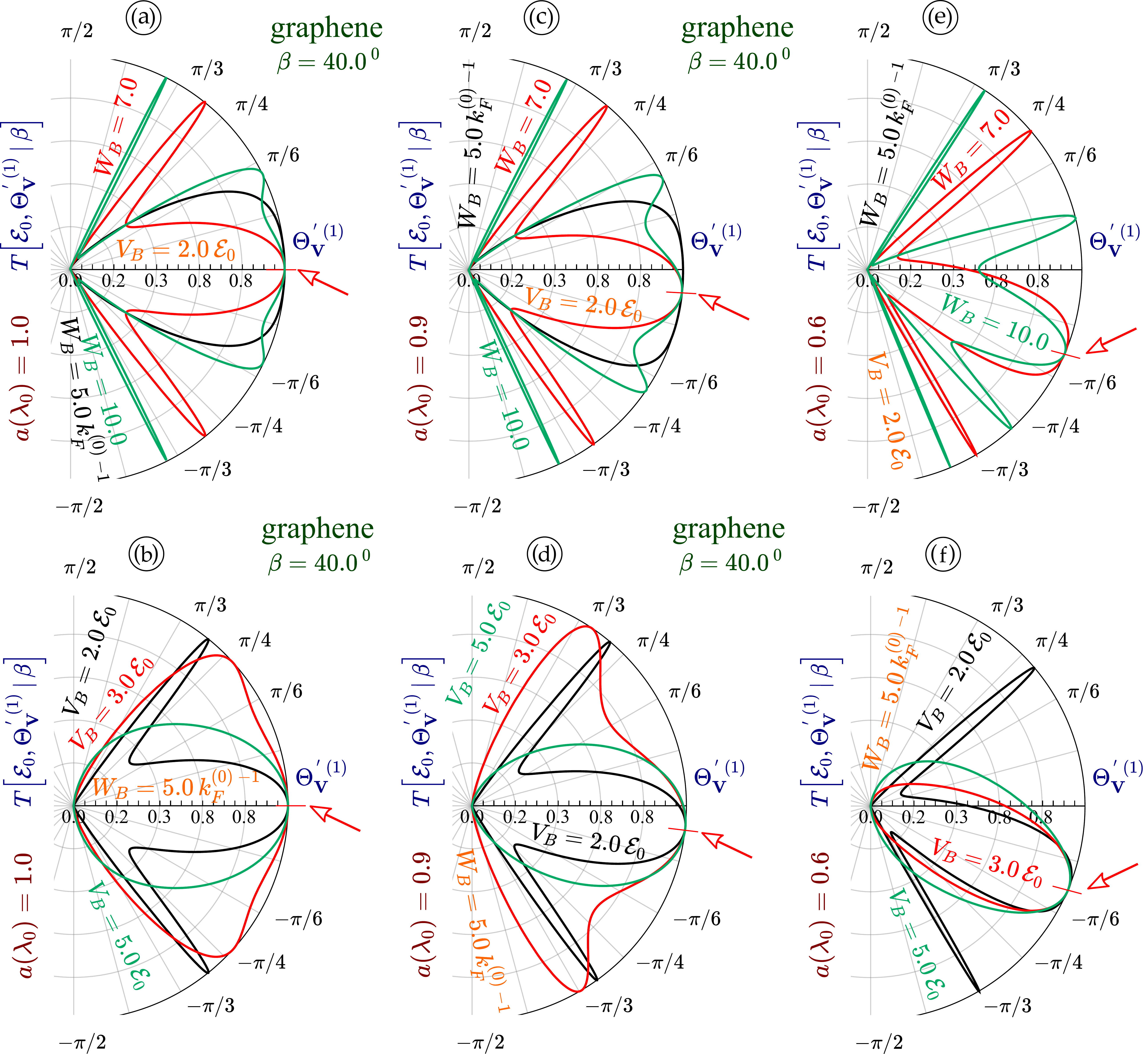}
\caption{(Color online) Angular plots for $T\left(\varepsilon_0, \Theta_{{\bf V}'}^{(1)}\, \vert \,\beta \right)$ 
as a function of $\Theta_{{\bf V}'}^{(1)}$ in graphene. 
Each panel relates to a specific value of $a_0(\lambda_0)=1.0$ for $(a)$, $(b)$; $0.9$ for $(c)$, $(d)$; $0.6$ for $(e)$, $(f)$.
Panels $(a)$, $(c)$, $(e)$ demonstrate the transmission by black, red and green curves for $k_F^{(0)}W_B = 5,\,7,\,10$ and $V_B/\epsilon_0=2$, as well as for 
$V_B/\varepsilon_0 = 2,\,3,\,5$ and $k_F^{(0)}W_B=5$ in $(b)$, $(d)$, $(f)$.
The direction of the shifted non-head-on Klein paradox is indicated by the red arrow in each panel. 
Here, $\beta=40^{\,\rm o}$ is set for all panels.}
\label{FIG:7}
\end{figure}

\begin{figure} 
\centering
\includegraphics[width=0.9\textwidth]{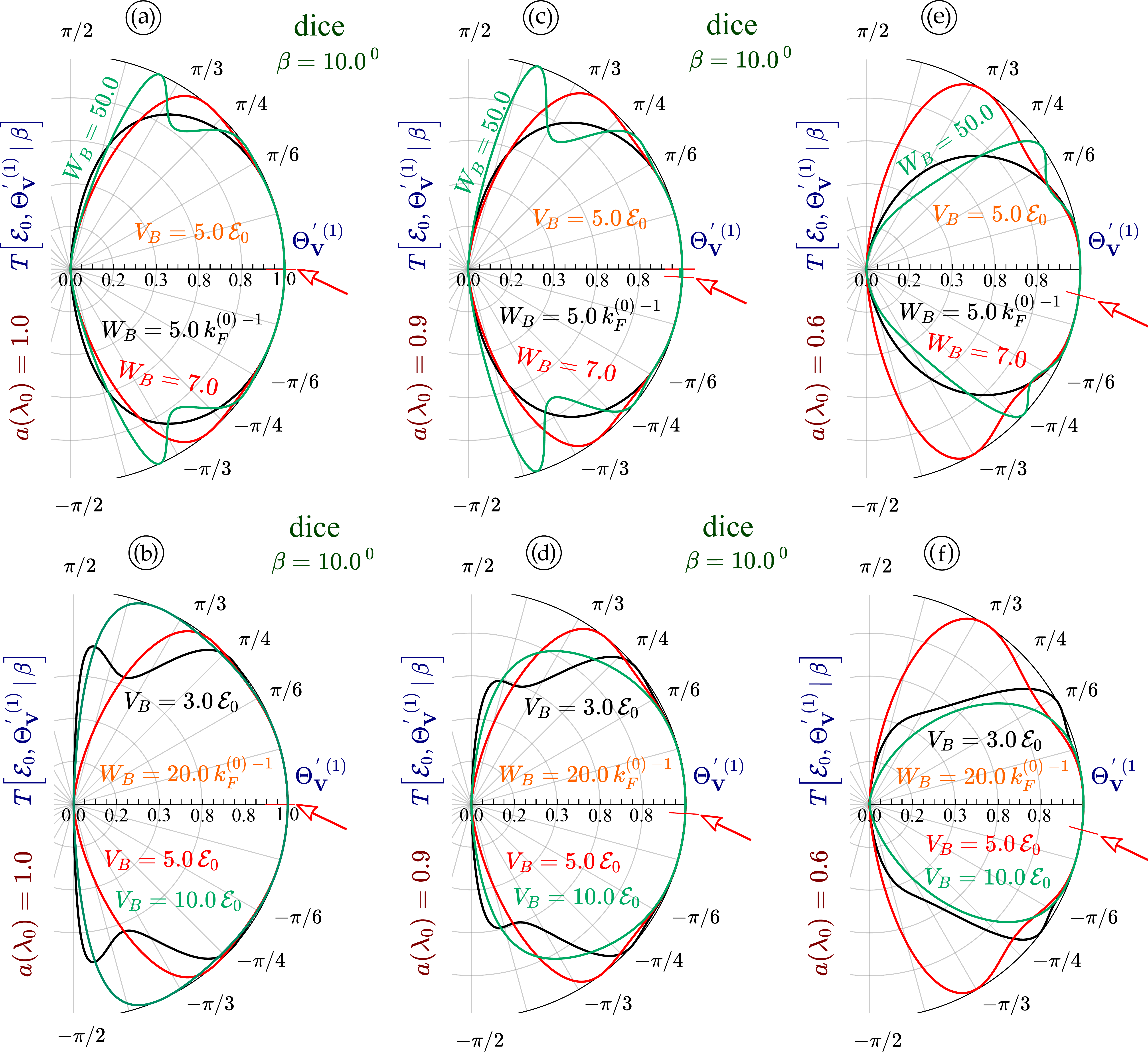}
\caption{(Color online) Angular plots for $T\left(\varepsilon_0, \Theta_{{\bf V}'}^{(1)}\, \vert \,\beta \right)$ 
as a function of $\Theta_{{\bf V}'}^{(1)}$ in dice lattices. 
Each panel relates to a specific value of $a_1(\lambda_0)=1.0$ for $(a)$, $(b)$; $0.9$ for $(c)$, $(d)$; $0.6$ for $(e)$, $(f)$.
Panels $(a)$, $(c)$, $(e)$ demonstrate the transmission by black, red and green curves for $k_F^{(0)}W_B = 5,\,7,\,50$ and $V_B/\varepsilon_0=5$, as well as for 
$V_B/\varepsilon_0 = 3,\,5,\,10$ and $k_F^{(0)}W_B=20$ in $(b)$, $(d)$, $(f)$.
The direction of the shifted non-head-on Klein paradox is indicated by the red arrow in each panel. 
Here, $\beta=10^{\,\rm o}$ is set for all panels.}		
\label{FIG:12}
\end{figure}

\begin{figure} 
\centering
\includegraphics[width=0.9\textwidth]{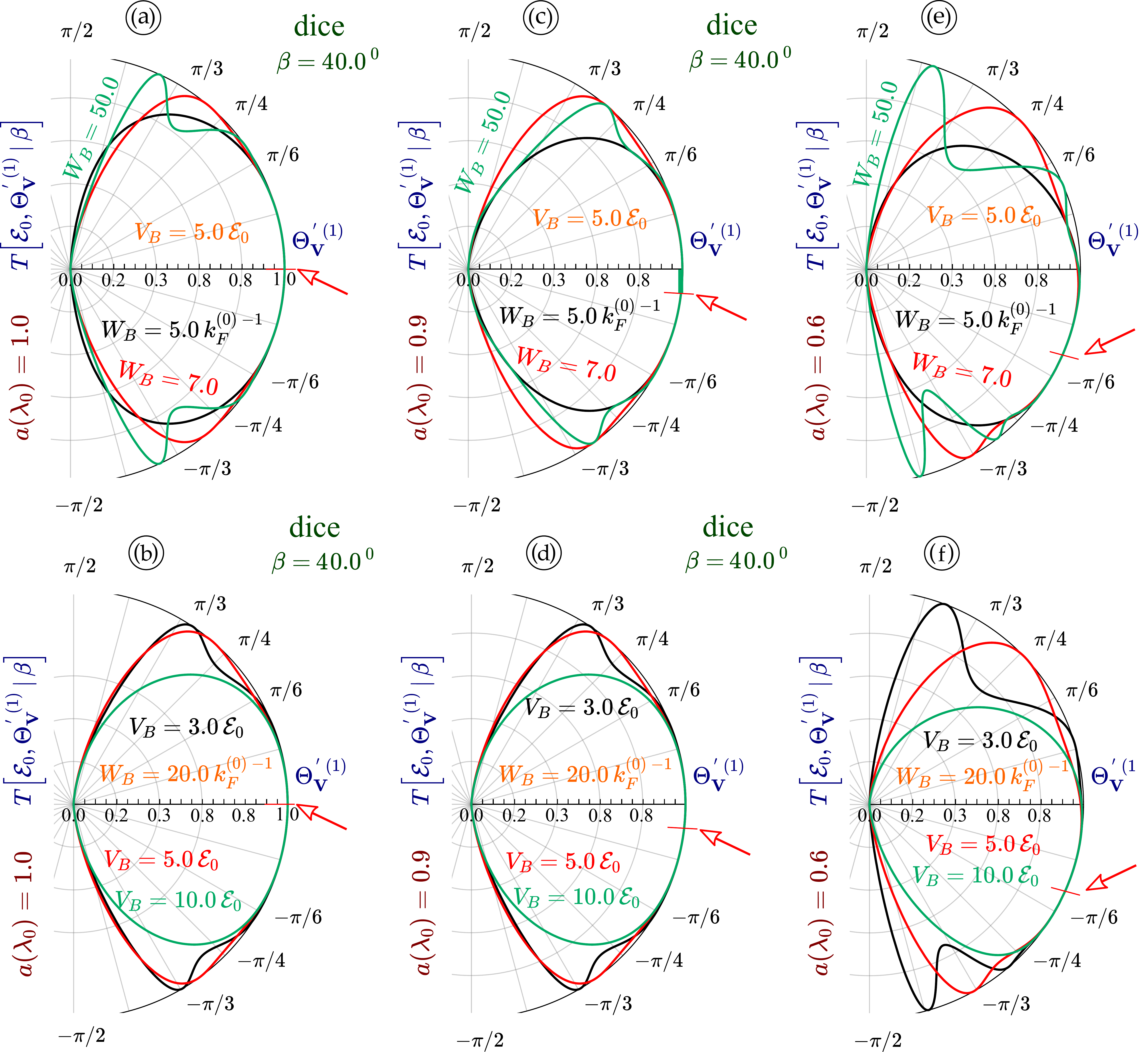}
\caption{(Color online) Angular plots for $T\left(\varepsilon_0, \Theta_{{\bf V}'}^{(1)}\, \vert \,\beta \right)$ 
as a function of $\Theta_{{\bf V}'}^{(1)}$ in dice lattices. 
Each panel relates to a specific value of $a_1(\lambda_0)=1.0$ for $(a)$, $(b)$; $0.9$ for $(c)$, $(d)$; $0.6$ for $(e)$, $(f)$.
Panels $(a)$, $(c)$, $(e)$ demonstrate the transmission by black, red and green curves for $k_F^{(0)}W_B = 5,\,7,\,50$ and $V_B/\varepsilon_0=5$, as well as for 
$V_B/\varepsilon_0 = 3,\,5,\,10$ and $k_F^{(0)}W_B=20$ in $(b)$, $(d)$, $(f)$.
The direction of the shifted non-head-on Klein paradox is indicated by the red arrow in each panel.
Here, $\beta=40^{\,\rm o}$ is set for all panels.}		
\label{FIG:13}
\end{figure}

\begin{figure} 
\centering
\includegraphics[width=0.9\textwidth]{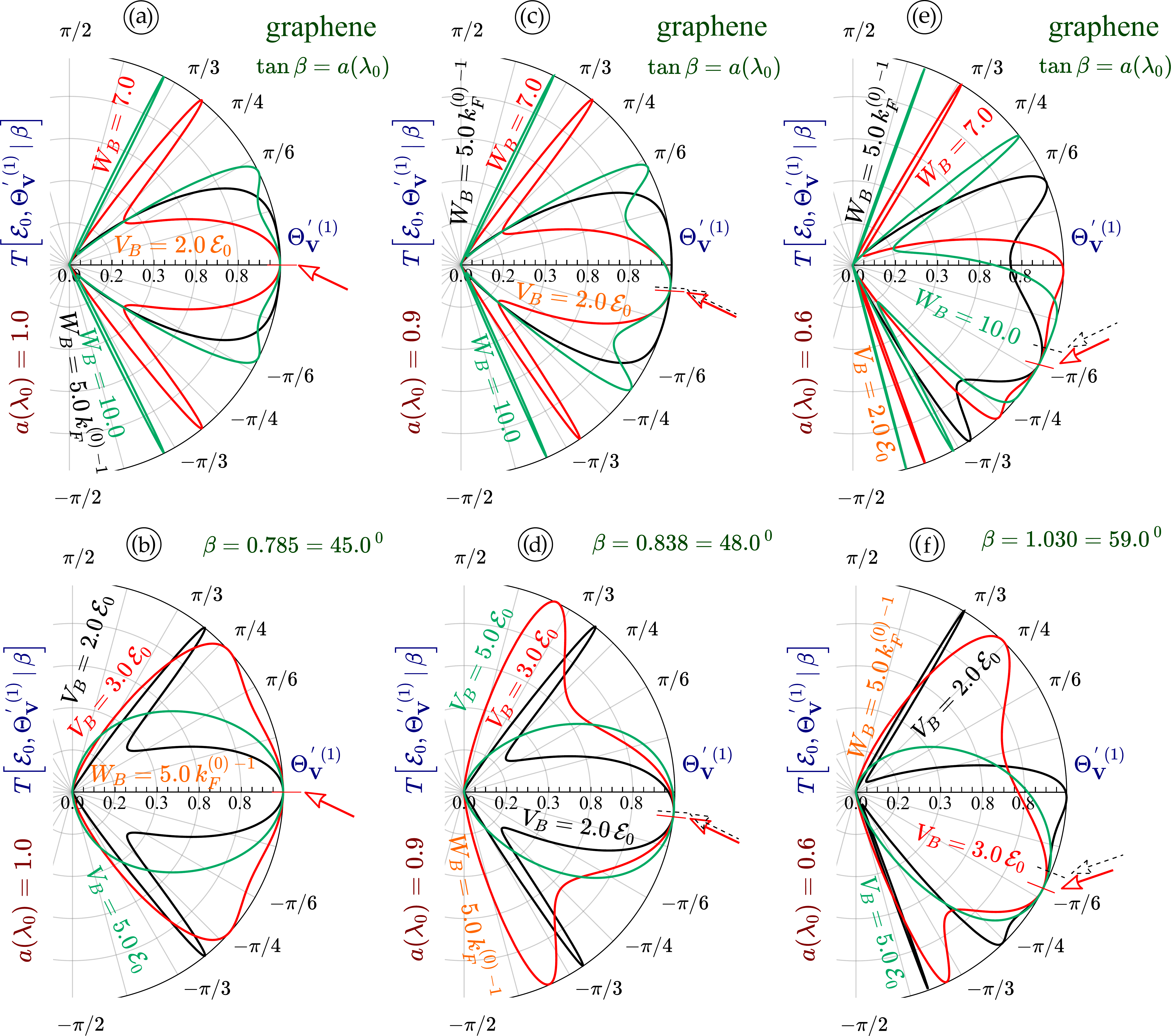}
\caption{(Color online) Angular plots for $T\left(\varepsilon_0, \Theta_{{\bf V}'}^{(1)}\, \vert \,\beta \right)$ 
as a function of $\Theta_{{\bf V}'}^{(1)}$ in graphene. 
Each panel relates to a specific value of $a_0(\lambda_0)=1.0$ for $(a)$, $(b)$; $0.9$ for $(c)$, $(d)$; $0.6$ for $(e)$, $(f)$.
Panels $(a)$, $(c)$, $(e)$ demonstrate the transmission by black, red and green curves for $k_F^{(0)}W_B = 5,\,7,\,10$ and $V_B/\varepsilon_0=2$, as well as for 
$V_B/\varepsilon_0 = 2,\,3,\,5$ and $k_F^{(0)}W_B=5$ in $(b)$, $(d)$, $(f)$.
The direction of the shifted non-head-on Klein paradox is indicated by the red arrow in each panel. 
Here, $\tan\beta_M=1/a_0(\lambda_0)$ is set for all panels, and then $\beta_M=45^{\,\rm o},\ 48^{\,\rm o},\ 59^{\,\rm o}$ correspond to $(a)$-$(b)$,  $(c)$-$(d)$, $(e)$-$(f)$, respectively.}
\label{FIG:8}
\end{figure}

\begin{figure} 
\centering
\includegraphics[width=0.9\textwidth]{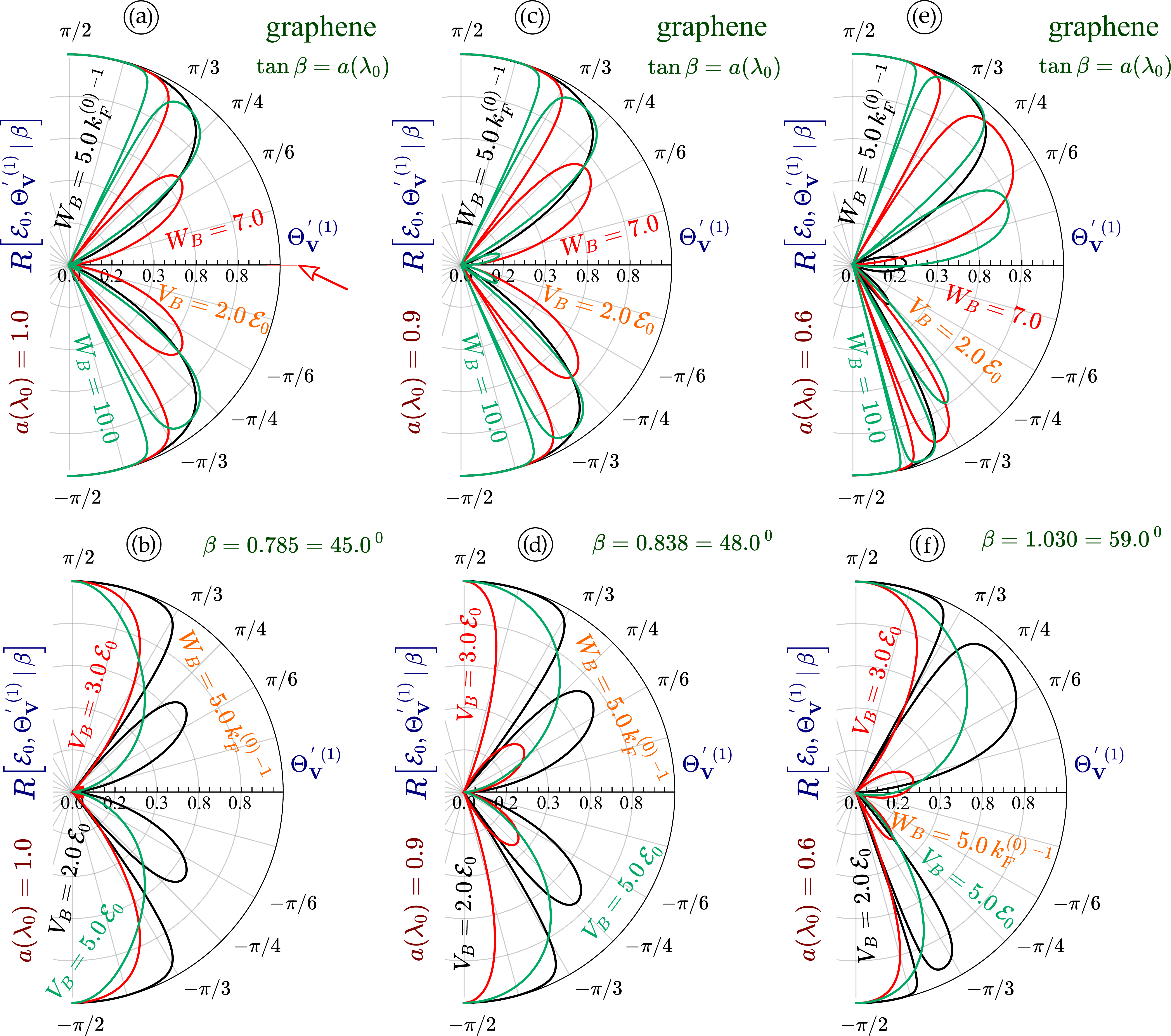}
\caption{(Color online) Angular plots for the reflection $R\left(\varepsilon_0, \Theta_{{\bf V}'}^{(1)}\, \vert \,\beta \right)$ 
as a function of $\Theta_{{\bf V}'}^{(1)}$ in graphene. 
Each panel relates to a specific value of $a_0(\lambda_0)=1.0$ for $(a)$, $(b)$; $0.9$ for $(c)$, $(d)$; $0.6$ for $(e)$, $(f)$.
Panels $(a)$, $(c)$, $(e)$ demonstrate the reflection by black, red and green curves for $k_F^{(0)}W_B = 5,\,7,\,10$ and $V_B/\varepsilon_0=2$, as well as for 
$V_B/\varepsilon_0 = 2,\,3,\,5$ and $k_F^{(0)}W_B=5$ in $(b)$, $(d)$, $(f)$.
Here, $\tan\beta_M=1/a_0(\lambda_0)$ is set for all panels, and then $\beta_M=45^{\,\rm o},\ 48^{\,\rm o},\ 59^{\,\rm o}$ correspond to $(a)$-$(b)$,  $(c)$-$(d)$, $(e)$-$(f)$, respectively.}
\label{FIG:9}
\end{figure}

\begin{figure} 
\centering
\includegraphics[width=0.9\textwidth]{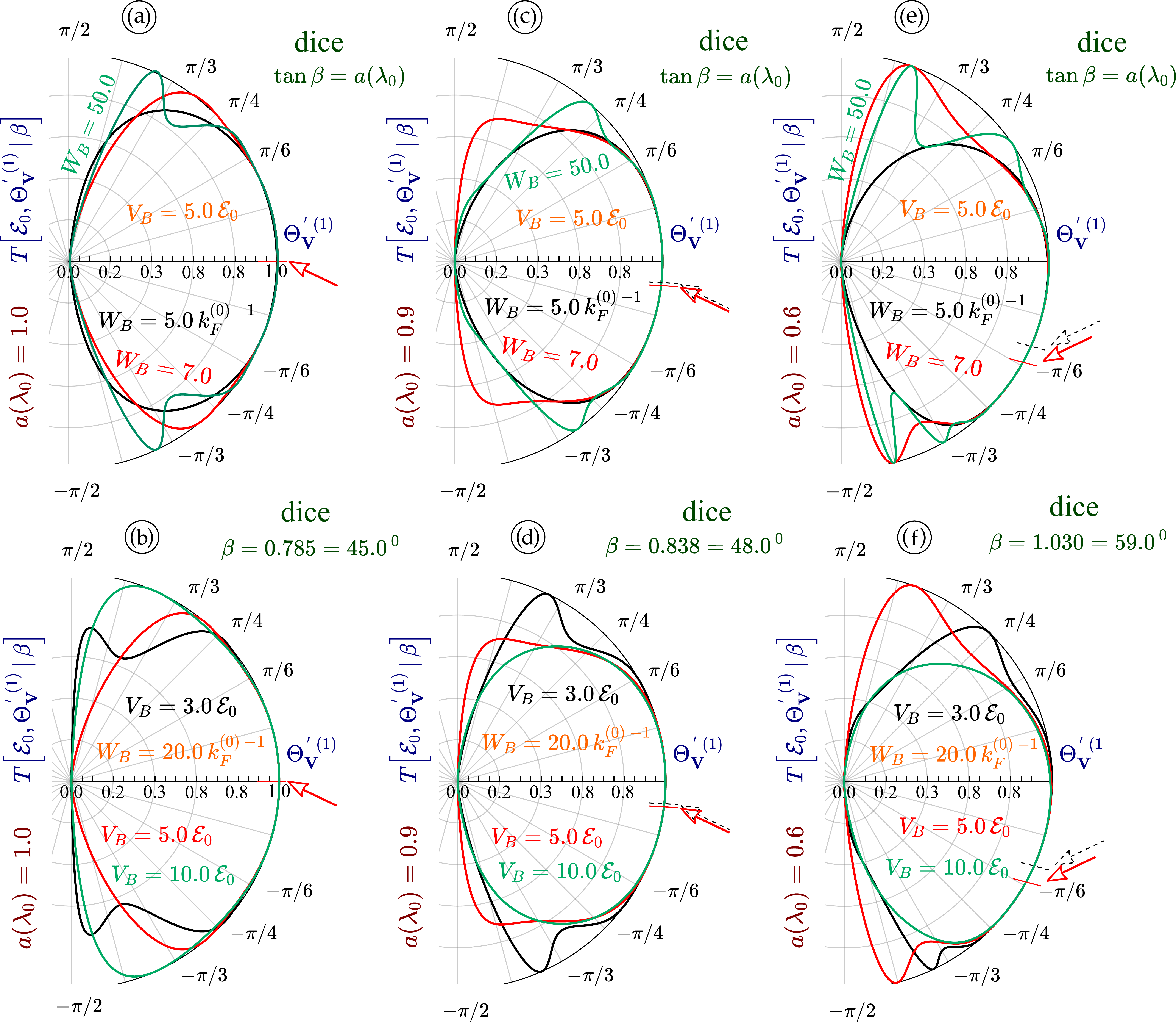}
\caption{(Color online) Angular plots for $T\left(\varepsilon_0, \Theta_{{\bf V}'}^{(1)}\, \vert \,\beta \right)$ 
as a function of $\Theta_{{\bf V}'}^{(1)}$ in dice lattices. Each panel relates to a specific value of $a_1(\lambda_0)
=1.0$ for $(a)$, $(b)$; $0.9$ for $(c)$, $(d)$; $0.6$ for $(e)$, $(f)$. Panels $(a)$, $(c)$, $(e)$ demonstrate the 
transmission by black, red and green curves for $k_F^{(0)}W_B = 5,\,7,\,50$ and $V_B/\varepsilon_0=5$, as well as for 
$V_B/\varepsilon_0 = 3,\,5,\,10$ and $k_F^{(0)}W_B=20$ in $(b)$, $(d)$, $(f)$. The direction of the shifted non-head-on
Klein paradox is indicated by the red arrow in each panel. Here, $\tan\beta_M=1/a_1(\lambda_0)$ is set for all panels, 
and then $\beta_M=45^{\,\rm o},\ 48^{\,\rm o},\ 59^{\,\rm o}$ correspond to $(a)$-$(b)$,  $(c)$-$(d)$, $(e)$-$(f)$, 
respectively.}
\label{FIG:14}
\end{figure}

\begin{figure} 
\centering
\includegraphics[width=0.9\textwidth]{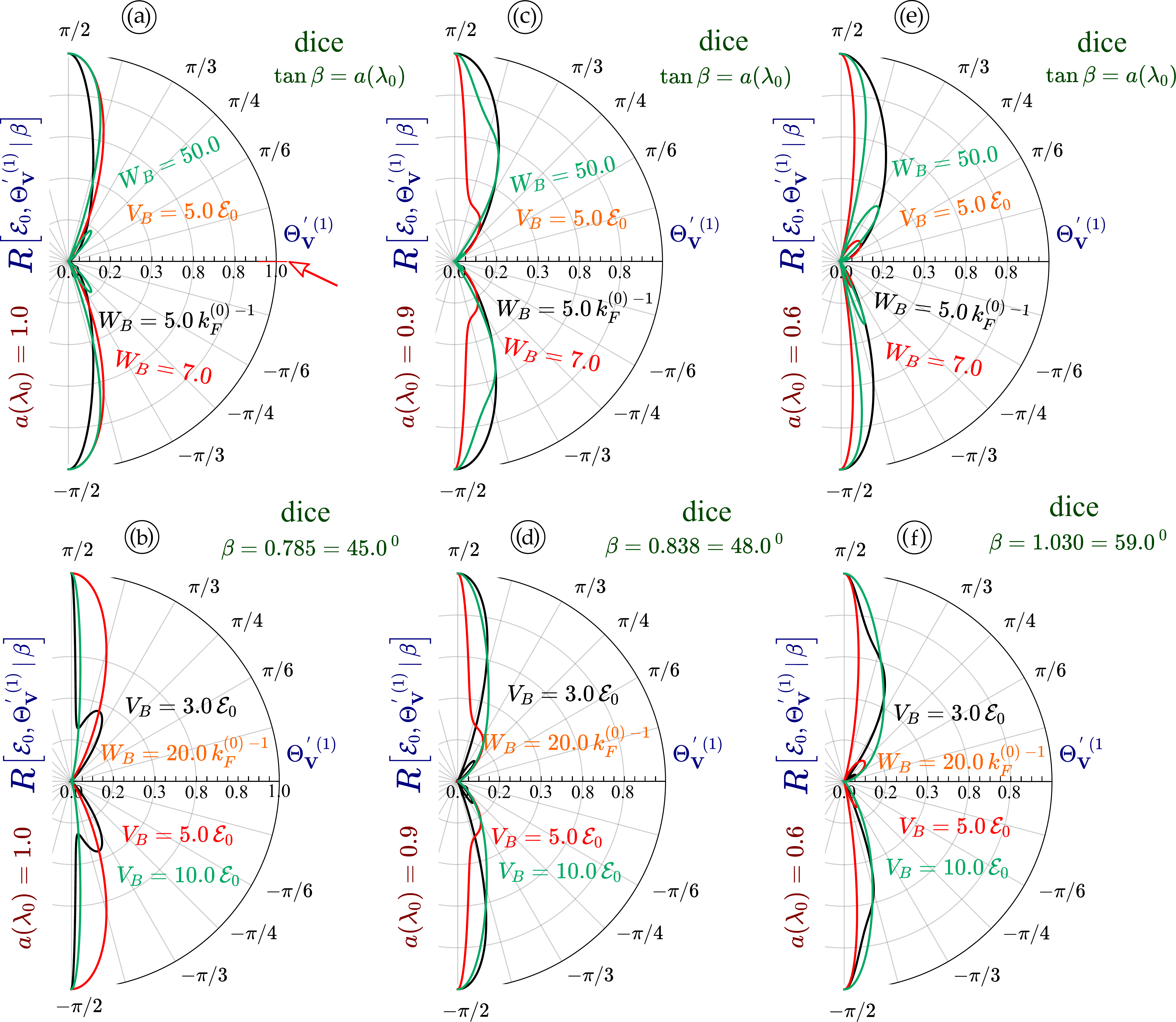}
\caption{(Color online) Angular plots for the reflection $R\left(\varepsilon_0, \Theta_{{\bf V}'}^{(1)}\, \vert \,\beta \right)$ 
as a function of $\Theta_{{\bf V}'}^{(1)}$ in dice lattices. 
Each panel relates to a specific value of $a_1(\lambda_0)=1.0$ for $(a)$, $(b)$; $0.9$ for $(c)$, $(d)$; $0.6$ for $(e)$, $(f)$.
Panels $(a)$, $(c)$, $(e)$ demonstrate the reflection by black, red and green curves for $k_F^{(0)}W_B = 5,\,7,\,50$ and $V_B/\varepsilon_0=5$, as well as for 
$V_B/\varepsilon_0 = 3,\,5,\,10$ and $k_F^{(0)}W_B=20$ in $(b)$, $(d)$, $(f)$.
Here, $\tan\beta_M=1/a_1(\lambda_0)$ is set for all panels, and then $\beta_M=45^{\,\rm o},\ 48^{\,\rm o},\ 59^{\,\rm o}$ correspond to $(a)$-$(b)$,  $(c)$-$(d)$, $(e)$-$(f)$, respectively.}
\label{FIG:15}
\end{figure}

\begin{figure} 
\centering
\includegraphics[width=0.55\textwidth]{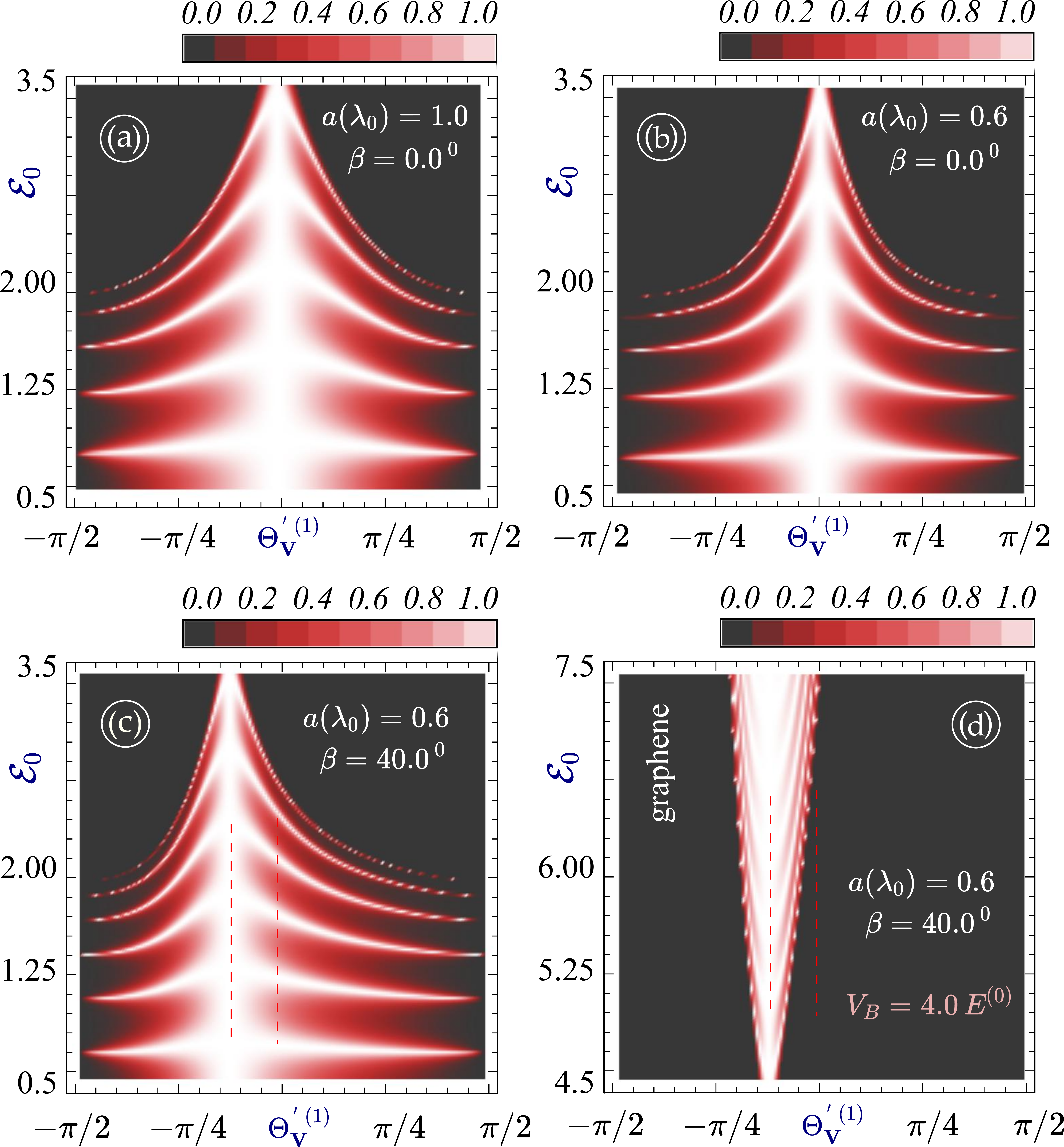}
\caption{(Color online) Density plots for $T\left(\varepsilon_0, \Theta_{{\bf V}'}^{(1)}\, \vert \,\beta \right)$ 
as functions of $\Theta_{{\bf V}'}^{(1)}$ and $\varepsilon_0/E_F^{(0)}$ in graphene with $V_B/E_F^{(0)}=4$. Panel $(a)$ is for 
$a_0(\lambda_0) = 1$, while panels $(b)$-$(d)$ are for $a_0(\lambda_0) = 0.6$. Upper plots $(a)$-$(b)$ 
correspond to $\beta = 0^{\,\rm o}$, but lower plots to $\beta= 40^{\,\rm o}$. Panels $(c)$-$(d)$ are plotted for the
same values of $a_0(\lambda_0)$ and $\beta$, and therefore, differ only by the energy range for display. Particularly, plot $(d)$ shows the transmission for 
the range of $\varepsilon_0$ above the barrier height $V_B$.}
\label{FIG:10}
\end{figure}

\begin{figure} 
\centering
\includegraphics[width=0.55\textwidth]{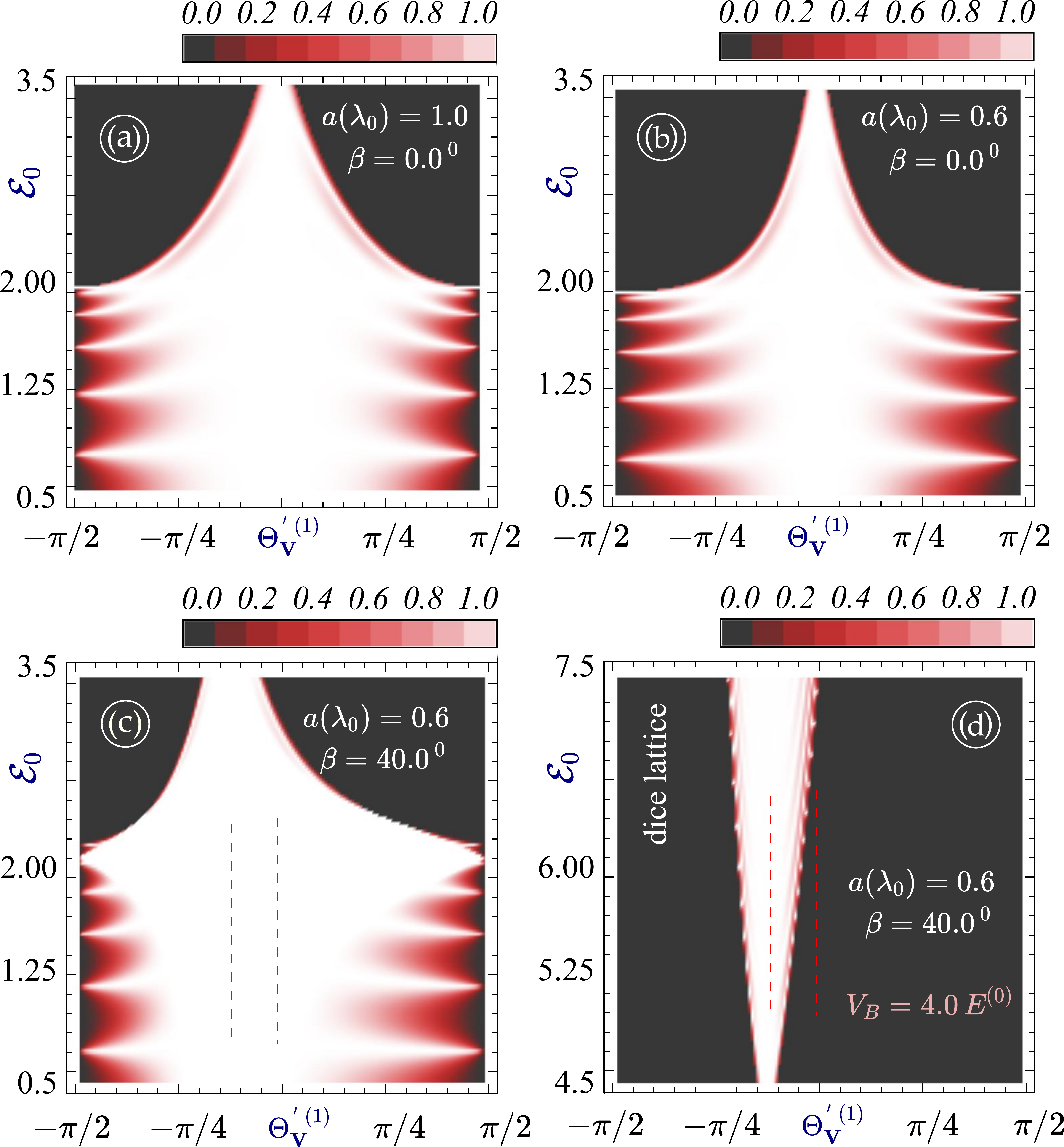}
\caption{(Color online) Density plots for $T\left(\varepsilon_0, \Theta_{{\bf V}'}^{(1)}\, \vert \,\beta \right)$ 
as functions of $\Theta_{{\bf V}'}^{(1)}$ and $\varepsilon_0/E_F^{(0)}$ in dice lattices with $V_B/E_F^{(0)}=4$. Panel $(a)$ is for 
$a_1(\lambda_0) = 1$, while panels $(b)$-$(d)$ are for $a_1(\lambda_0) = 0.6$. Upper plots $(a)$-$(b)$ 
correspond to $\beta = 0^{\,\rm o}$, but lower plots to $\beta= 40^{\,\rm o}$. Panels $(c)$-$(d)$ are plotted for the
same values of $a_1(\lambda_0)$ and $\beta$, and therefore, differ only by the energy range for display. Particularly, plot $(d)$ shows the transmission for 
the range of $\varepsilon_0$ above the barrier height $V_B$.}
\label{FIG:16}
\end{figure}

In the case of the anomalous tunneling, it is rather hard to determine which peak of the electron transmission is associated with Klein paradox, 
while all the other peaks represent different types of transmission resonances and are out of the focus of our current study. 
For this reason, we always display transmission results with different widths and heights of potential barrier in each polar plot, as shown in 
Figs.\,\ref{FIG:5}-\ref{FIG:10} for graphene and in Figs.\,\ref{FIG:11}-\ref{FIG:16} for a dice lattice, respectively.
In this way, the position of the anomalous Klein paradox can be unambiguously determined. 
\medskip

Technically, the anisotropy in the electron dispersions could be tuned externally by varying the intensity of applied linearly-polarized irradiation 
or the electron-light coupling $\lambda_0$. Apart from that, the angle $\beta$ between the surface normal of barrier in the $\{x',y'\}$ frame 
and the long-axis of elliptical energy dispersion in the $\{x,y\}$ frame 
could vary from zero up to nearly $\pi/2$ together with the polarization angle $\beta_0$ for the imposed dressing field. 
We expect that all of these factors can greatly affect the anomalous Klein tunneling. 
\medskip

We begin with the comparison having $\beta = 0^{\,\rm o}$, as shown in Fig.\,\ref{FIG:5} for graphene and in Fig.\,\ref{FIG:11} for a dice lattice. 
For graphene in Fig.\,\ref{FIG:5}, we see that for all the cases with any anisotropy but no rotation between two frames, the direction of the anomalous Klein paradox remains identical to that 
for a head-on incidence $\Theta_{{\bf V}'}^{(1)} = 0$. However, a finite anisotropy still affects the transmission greatly by narrowing its peak
and also the angle distribution range, as seen from panels $(a)$, $(c)$, $(e)$ of Fig.\,\ref{FIG:5}. Meanwhile, we find that
the shapes of transmission peaks off the $\Theta_{{\bf V}'}^{(1)} = 0$ axis depend sensitively on the barrier width $W_B$ for panels $(a)$, $(c)$, $(e)$ or barrier height $V_B$ 
for panels $(b$), $(d)$, $(f)$ 
in analogy to that in Eq.\,\eqref{tneto} for an isotropic 
Dirac cone, where the number of oscillations in $\cos^2(k_x^{(2)} W_B)$ increases with $W_B$. 
For dice lattices, on the other hand, a very broad peak for the anomalous Klein paradox is found in comparison with graphene. Moreover, the variations in angle distribution of transmission 
are only limited to a large-angle range for the individual change of $W_B$, $V_B$ and $a_1(\lambda_0)$.
\medskip

Next, we introduce a misalignment angle $\beta$ between the electron wave numbers $k_x$ and $k_{x'}$ and compare the results in Figs.\,\ref{FIG:6} and \ref{FIG:7} for graphene 
and in Figs.\,\ref{FIG:12} and \ref{FIG:13} for dice lattices with $\beta=10^{\,\rm o}$ and $40^{\,\rm o}$, respectively. 
For isotropic energy dispersion with $a_0(\lambda_0)=1$, the direction of the Klein paradox always remains at the angle of $\Theta_{{\bf V}'}^{(1)}=0$, independent of $\beta$.
For graphene in Figs.\,\ref{FIG:6} and \ref{FIG:7}, we find that the direction of the anomalous Klein paradox gradually moves downward away from the angle $\Theta_{{\bf V}'}^{(1)}=0$ 
with increasing $\beta$ from zero and as $a_0(\lambda_0)$ is reduced from unity for enhanced anisotropy in energy dispersions of electrons, as seen from $(e)$ and $(f)$ of Fig.\,\ref{FIG:6}. 
Moreover, such a unique feature is further enforced due to increased $\beta$ by comparing panels $(e)$ and $(f)$ of Fig.\,\ref{FIG:6} with these two panels of Fig.\,\ref{FIG:7}.
For dice lattices in Figs.\,\ref{FIG:12}-\ref{FIG:13}, on the other hand, the deviation from the angle $\Theta_{{\bf V}'}^{(1)}=0$ with enhanced anisotropy in energy dispersions by reduced
$a_1(\lambda_0)$ becomes less evident due to a very broad anomalous-Klein-paradox peak in this case, as shown in panels $(e)$ and $(f)$ of Figs.\,\ref{FIG:12} and \ref{FIG:13}.
However, the angle deviation from $\Theta_{{\bf V}'}^{(1)}=0$ still increases with the misalighment angle $\beta$ for dice lattices.
Moreover, the direction of the anomalous-Klein-paradox peak is found fixed in both Figs.\,\ref{FIG:6} and \ref{FIG:7} and Figs.\,\ref{FIG:12} and \ref{FIG:13},
although the angle distributions of other resonant-tunneling peaks change with either barrier width $W_B$ or barrier height $V_B$.
\medskip

Furthermore, we consider the transmission and reflection with a specific rotation angle $\beta_M = \tan^{-1}[1/a_{\{0,1\}}(\lambda_0)]$ in Figs.\,\ref{FIG:8} and \ref{FIG:9} for graphene 
and in Figs.\,\ref{FIG:14} and \ref{FIG:15} for dice lattices, which leads to the 
biggest angle deviation between the anomalous Klein tunneling direction and the direction of head-on incidence. 
The result for the electron transmission are presented separately in Figs.\,\ref{FIG:8} and \ref{FIG:14} for graphene and dice lattices. 
Indeed, we find that the angle for the anomalous Klein tunneling increases with reducing $a_{\{0,1\}}(\lambda_0)$ value, 
but the resulting variation becomes noticeable only for a larger anisotropy with $a_{\{0,1\}}(\lambda_0) = 0.6$. However, the condition for $a_{\{0,1\}}(\lambda_0) = 0.6$  
cannot be met by applying an off-resonance dressing field, and therefore, the results presented in Figs.\,\ref{FIG:8} and \ref{FIG:14} are only for the comparison purpose. 
On the other hand, the reflection graphs in Fig.\,\ref{FIG:9} for graphene and in Fig.\,\ref{FIG:15} for dice lattices are just used to confirm and uphold 
our previous calculations\,\cite{oura} since the maximum of the transmission should correlate to the independently-calculated vanishing reflection, as can be verified from 
Figs.\,\ref{FIG:8}-\ref{FIG:15}. 
\medskip 
   
In spite of the fact that only the derived boundary conditions for a dice lattice are new ingredients in this paper in comparison with the boundary conditions for graphene, 
to serve the purpose of comparison, the numerical results about the anomalous 
Klein tunneling with $a_{\{0,1\}}(\lambda_0)\neq 1$ 
have been presented for both cases, where the incident kinetic energy $\varepsilon_0$ of electrons, as well as the angle $\beta$ between the longitudinal wave numbers 
$k_x$ in the $\{x,y\}$ frame for the energy dispersion and $k_{x'}$ in the  $\{x',y'\}$ frame for the surface normal of potential barrier,
are assumed the same for both materials. In general, we expect that the transmission for a dice lattice is considerably larger than that for graphene under similar
conditions, as seen especially well from the density plots in Fig.\,\ref{FIG:16} for a significntly expanded white region.
\medskip

The magic case for a complete transmission covering the full range of incident angles $\Theta_{{\bf V}'}^{(1)}$ if the incoming particle energy is $\varepsilon_0/E_F^{(0)}=1/2$, 
remains in place for anisotropic dispersions and a finite rotation angle $\beta \neq 0$. Figure\ \ref{FIG:10} for graphene and Fig.\,\ref{FIG:16} for dice lattices
demonstrate clearly that the direction of the anomalous Klein paradox does not depend on the kinetic energy $\varepsilon_0$ of 
incoming particles, which is in agreement with our theoretical model. Meanwhile, the transmission results in Fig.\,\ref{FIG:16} for dice lattices do not
display any dependence on the valley index $\tau = \pm 1$ even though $\tau$ appears in the boundary conditions in Eq.\,\eqref{MainSystD2}. However, this conclusion is not
expected to be the case for general $\alpha-\mc{T}_3$ lattices.          

\clearpage
\section{Conclusions and Discussions} 
\label{sec5}

In conclusion, we have studied in this paper tunneling related to the anomalous Klein paradox, which results uniquely from the misalignment of optically-controlled elliptical dispersion 
for Dirac-cone dressed states and the direction of incident kinetic particles in our considered system. Specifically,
we have performed a thorough theoretical investigation on the Klein tunneling over a square finite-width potential barrier 
in graphene and a pseudospin-$1$ dice lattice with laser-induced anisotropic energy dispersions in their valence and conduction bands. Such 
a tunable anisotropy can be prepared by applying a linearly-polarized off-resonance dressing field with the polarization
direction away from the long-axis of elliptical energy dispersion of Dirac-cone dressed states.  
\medskip

The electron dynamics of optically-controllable dressed states have been explored theoretically 
by using Floquet-Magnus perturbative-expansion approach for electron-light interaction Hamiltonian.
In particular, we find that the effect of a high-frequency linearly-polarized irradiation for both graphene and a dice lattice can lead 
to an intensity-dependent modification to the quantum phases of dressed states, which is characterized by a spinor angle different from the incident angle of incoming particles.
Here, the direction of incident particles is measured against the surface normal of a potential barrier, while the direction of the spinor vector of a 
dressed-state wave-function is quantified with respect to 
the longer axis of the elliptical energy dispersion of electrons. Therefore, two individual coordinate frames must be introduced specifically for these two different directions. 
Physically, however, there exists a angle misalignment between these two frames due to the presence of intensity-dependent anisotropy in energy dispersion 
of dressed-state electrons under a linearly-polarized irradiation.
\medskip

Using unique electronic properties from these linear-polarization coupled electron dressed states in graphene and dice lattices, we have 
investigated the their transmission and found the appearance of the so-called anomalous Klein paradox with a peak in the angle distribution of transmib00ssion away from the 
head-on direction for incoming particles.
This resulting finite incident angle depends on the degree of anisotropic energy dispersion $a_{\{0,1\}}(\lambda_0)$ or the electron-light coupling constant $\lambda_0$, 
as well as on the misalignment angle $\beta$ between the surface normal of the potential barrier and the longer axis of the elliptical energy dispersion of the dressed-state electrons.
Moreover, the maximum angle deviation for the anomalous Klein paradox 
is achieved as $\beta=\beta_M\equiv\tan^{-1}[1/a_{\{0,1\}}(\lambda_0)]$, somewhat similar to the case of phosphorene\,\cite{LiuM} with material-based anisotropic band structures.
\medskip

Apart from the angle position of the anomalous Klein paradox, the angle distribution of other resonant peaks in both transmission and reflection appear quite different for graphene and  
dice lattices. 
Quantitatively, a dice lattice can acquire much larger off-peak transmission amplitudes compared to graphene under the same conditions, and in particular,
the ``magic case'' for a complete transmission covering the full range of incident angle is seen at $\varepsilon_0 =V_B/8$ 
for both graphene and dice lattices, independent of the degree of the anisotropy $a_{\{0,1\}}(\lambda_0)$ and the value of the misalignment angle $\beta$. 
\medskip

In the presence of a potential barrier, 
compared with the known boundary conditions for graphene with the pseudospin-$1/2$ Hamiltian in Eq.\,\eqref{graph}, our derived boundary conditions for a dice lattice 
with the pseudospin-$1$ Hamiltian acquire an addition constraint in Eq.\,\eqref{dice} and become quite different. 
These new boundary conditions can be employed 
for the calculation of electron transmission in an anisotropic dice lattice, such as an irradiated 
$SrTiO_3/SrIrO_3/SrTiO_3$ trilayer. 
\medskip

From an application perspective, our theoretical results could be practically implemented into extremely wide range of 
recently discovered Dirac materials either with a built-in anisotropic energy dispersion or with an externally-tunable anisotropy due to incident irradiation. 
In the absence of anisotropy in energy dispersion, our system behaviors much like $n$-$p$-$n$ multi-junctions with an additional electric gate 
to control an electrically-injected current by a positive base voltage for tuning barrier height $V_B$.
In the presence of laser-tunable anisotropic energy dispersion, on the other hand, an antenna-coupld incident laser can be employed as a laser-based gate 
to control both the magnitude and direction of an injected ballistic current through the angle-dependent electron transmission $T\left(\varepsilon_0, \Theta_{{\bf V}'}^{(1)}\, \vert \,\beta \right)$.
Undoubtedly, our explored and demonstrated properties for both coherent tunneling and ballistic transport of electrons will find their applications 
in constructing novel optical and electronic nano-scale switching devices.   

\section*{Acknowledgement(s)}
A.I. would like to acknowledge the funding provided by TRADA-51-82 PSC-CUNY Award \# 63061-00--51.
G.G. appreciates financial support from the Air Force Research Laboratory (AFRL) through 
grant FA9453-18-1-0100 and award FA2386-18-1-0120. D.H. thanks the supports from the Laboratory University
Collaboration Initiative (LUCI) program and from the Air Force Office of Scientific Research (AFOSR).

\clearpage
\appendix
\section{Pseudospin-1 $\alpha$-dependent Pauli matrices}
\label{apa}

Our pseudospin-$1$ Hamiltonian in Eq.\,\eqref{HamG} for arbitrary $\alpha = \tan \phi$ is defined in terms of the following two
$\phi$-dependent $3 \times 3$ matrices:

\begin{equation}
\label{Sxp}
\hat{\Sigma}_{x}^{(\alpha)} = \left[
\begin{array}{ccc}
 0 & \cos \phi & 0 \\
 \cos \phi & 0 & \sin \phi \\
 0 & \sin \phi & 0
\end{array}
\right] \ ,
\end{equation}

\begin{equation}
\label{Syp}
\hat{\Sigma}_{y}^{(\alpha)} = i \,\left[
\begin{array}{ccc}
 0 & -\cos \phi & 0 \\
 \cos \phi & 0 & -\sin \phi \\
 0 & \sin \phi & 0
\end{array}
\right] \ .
\end{equation}
In this paper, we focus on a dice lattice with $\phi=\pi/4$, so that the matrices in Eqs.\,\eqref{Sxp} and \eqref{Syp} 
reduce to the regular $3 \times 3$ Pauli matrices:
 
\begin{equation}
\hat{\Sigma}^{(1)}_{x} = \frac{1}{\sqrt{2}} \, \left[
\begin{array}{ccc}
 0 & 1 & 0 \\
 1 & 0 & 1 \\
 0 & 1 & 0
\end{array}
\right] \ ,
\label{sig1}
\end{equation}

\begin{equation}
\hat{\Sigma}^{(1)}_{y} = \frac{i}{\sqrt{2}} \, \left[
\begin{array}{ccc}
 0 & -1 & 0 \\
 1 & 0 & -1 \\
 0 & 1 & 0
\end{array}
\right] \ .
\label{sig2}
\end{equation}
Sometimes, the third Pauli matrix

\begin{equation}
\hat{\Sigma}^{(1)}_{z} = \left[
\begin{array}{ccc}
 1 & 0 & 0 \\
 0 & 0 & 0 \\
 0 & 0 & -1
\end{array}
\right] \, 
\end{equation}
is also employed to introduce an energy gap to a pseudospin-$1$ Hamiltonian. 
\medskip

Since all our matrices, including the additional interacting Hamiltonian terms derived in the next Appendix\ \ref{apb}, are
Hermitian, it is convenient to introduce two new matrices

\begin{equation}
\hat{\Sigma}^{(1)}_\pm = \frac{1}{\sqrt{2}} \left(\hat{\Sigma}^{(1)}_x \pm i \, \hat{\Sigma}^{(1)}_y\,\right) \ ,
\end{equation}   
which have the following structure:

\begin{equation}
\hat{\Sigma}^{(1)}_+ =
\left[ 
\begin{array}{cc} 
\begin{array}{c} 
0 \\ 0 
\end{array}   
& \text{\Large $\hat{\mbb{I}}_{2}$} \\
0 & \begin{array}{cc}
0 & 0 
\end{array}
\end{array}
\right]
 =
\left[
\begin{array}{ccc}
 0 & 1 & 0 \\
 0 & 0 & 1 \\
 0 & 0 & 0
\end{array}
\right] 
\end{equation}

\begin{equation}
\hat{\Sigma}^{(1)}_- =
\left[ 
\begin{array}{cc}     
\begin{array}{cc}
0 & 0 
\end{array}
& 0  \\    
\text{\Large $\hat{\mbb{I}}_{2}$} &     
\begin{array}{c} 
0 \\ 0 
\end{array}   
\end{array}
\right] =
\left[
\begin{array}{ccc}
0 & 0 & 0 \\
1 & 0 & 0 \\
0 & 1 & 0
\end{array}
\right] \, ,
\end{equation}
where $\hat{\mbb{I}}_{2}$ is a regular $2 \times 2$ unit matrix corresponding to pseudospin-$1/2$ system, i.e.,

\begin{equation}
\hat{\mbb{I}}_{2} = \left[ 
\begin{array}{cc}
1 & 0 \\
0 & 1
\end{array}
\right] \ .
\end{equation}
Importantly, the obtained matrices $\hat{\Sigma}^{(1)}_\pm$ satisfy the simple commutation relations:  $\left[ \hat{\Sigma}^{(1)}_{+}, \hat{\Sigma}^{(1)}_{-} \right] = 2\,\hat{\Sigma}^{(1)}_{z}$ and 
$\left[ \hat{\Sigma}^{(1)}_{z}, \hat{\Sigma}^{(1)}_{\pm} \right] = 2\,\hat{\Sigma}^{(1)}_{\pm}$. 

\section{Hamiltonian of Electrons in a Dice Lattice with Arbitrary Direction of Light Polarization} 
\label{apb}

In contrast to our derived Eq.\,\eqref{t20}, we now consider an arbitrary direction for light polarization. We aim to obtain 
the dressed states of electrons in a dice lattice under linearly-polarized light with the vector 
potential

\begin{equation}
\label{alinA}
\mbox{\boldmath$A$}^{(L)}(\beta_0, t) = 
\left[ \begin{array}{c}
A^{(L)}_x (\beta_0, t) \\
A^{(L)}_y (\beta_0, t)
\end{array}
\right] = \frac{\mc{E}_0}{\omega} \, \cos (\omega t )  \left[
\begin{array}{c}
\cos \beta_0\\
\sin \beta_0
\end{array}
\right]\ .
\end{equation}
As a result, the case for the $x$-direction light polarization is got simply by setting $\beta_0 = 0$. The new Hamiltonian is acquired
by the standard substitution of $k_{x,y} \rightarrow k_{x,y} - (e/\hbar)\, A^{(L)}_{x,y}(\beta_0, t)$ in the Hamiltonian 
for both components of the vector potential $\mbox{\boldmath$A$}^{(L)}(\beta_0, t)$.
\medskip

Since the Hamiltonian in the absence of irritation is linear in wave vector $\mbox{\boldmath$k$}$ for electrons, the effect of imposed irradiation can be included by  

\begin{equation}
\label{aTlinham}
\hat{\mc{H}}_{1,\tau}(\mbox{\boldmath$k$}) \Longrightarrow \hat{\mbb{H}}^{(L)}(\mbox{\boldmath$k$}, t \, \vert \, \beta) = 
\hat{\mc{H}}_{1,\tau}(\mbox{\boldmath$k$}) + \hat{\mc{H}}_I^{(L)}(\beta_0, t) \ , 
\end{equation}
where subscript-index ``$1$'' is associated with $\alpha = 1$ for a dice lattice.
Moreover, the interaction Hamiltonian term in Eq.\,\eqref{aTlinham} is given by

\begin{equation}
\label{aHAL}
\mc{\mbb{H}}_I^{(L)}(\beta_0,t) = - \frac{\tau c_0}{\sqrt{2}}\,  \cos (\omega t) \left[
\begin{array}{ccc}
0 & \tet{e}^{- i \tau \beta_0} & 0 \\
\tet{e}^{i \tau \beta_0}  & 0 & \tet{e}^{- i \tau \beta_0} \\
0 & \tet{e}^{i \tau \beta_0} & 0 
\end{array}
\right] \ ,
\end{equation}
where $\tau=\pm 1$ is the valley index, and 
the coupling constant $c_0 = ev_F{\cal E}_0/\omega$ is the same for all type of light polarizations, which implies that the polarization
effect on the energy dispersion becomes similar in magnitude but different in features.
\medskip

The periodic time dependence of the interaction Hamiltonian term $\hat{\mbb{H}}_A^{(L)}(\beta_0,t)$ in Eq.\,\eqref{aHAL} could be cast into the following form 

\begin{equation}
\hat{\mbb{H}}_I^{\,(L)}(\beta_0,t)  = \hat{\mbb{O}}_{1,\tau}(\beta_0) \, \tet{e}^{i \omega t} + 
\hat{\mbb{O}}^{\dagger}_{1,\tau}(\beta_0) \, \tet{e}^{-i \omega t} \ , 
\label{ooper}
\end{equation}
where the operator $\hat{\mbb{O}}_{1, \tau}(\beta_0)$ and its conjugate $\hat{\mbb{O}}^\dag_{1, \tau}(\beta_0)$ are time independent. 
It is straightforward to find the operator $\hat{\mbb{O}}_{1,\tau}(\beta_0)$ from Eq.\,\eqref{aHAL}, yielding 

\begin{equation}
\label{aOop}
\hat{\mbb{O}}_{1,\tau}(\beta_0) = - \frac{\tau c_0}{2\sqrt{2}}\left[ 
\begin{array}{ccc}
0 & \tet{e}^{- i \tau \beta_0} & 0 \\
\tet{e}^{i \tau \beta_0}  & 0 & \tet{e}^{- i \tau \beta_0} \\
0 & \tet{e}^{i \tau \beta_0} & 0 
\end{array}
\right]  \ , 
\end{equation}
and it is equivalent to Eq.\,\eqref{aHAL} except for the prefactor. Moreover, matrix $\hat{\mbb{O}}_{\, 1, \tau}(\beta_0)$ itself is Hermitian which is unique
for the linearly-polarized light and is not the case for any other types of elliptical polarization including the circular one. 
\medskip 

By using Eqs.\,\eqref{aTlinham} and \eqref{ooper}, the effective time-independent Hamiltonian can be derived based on the standard Floquet-Magnus expansion approach, given by

\begin{equation}
\label{aTmexp}
 \hat{\mc{H}}_{\text{eff}} = \hat{\mc{H}}_{1, \tau}(\mbox{\boldmath$k$}) + \frac{1}{\hbar \omega} \, \left[ 
\, \hat{\mbb{O}}_{1, \tau}, \, \hat{\mbb{O}}_{1, \tau}^{\dag} \,
 \right] + \frac{1}{2 (\hbar \omega)^2} \left\{
 \left[ \left[
 \, \hat{\mbb{O}}_{1, \tau}, \, \hat{\mc{H}}_{1, \tau}(\mbox{\boldmath$k$}) \, 
 \right], \, 
 \hat{\mbb{O}}_{1, \tau}^\dagger \,  
 \right]
 \,\, + \,\, h.c.
 \right\} \,\, + \cdots\ \ ,
\end{equation}
where the first term in the expansion is just the non-interacting Hamiltonian, while the second
term $\left[\hat{\mbb{O}}_{1, \tau}, \, \hat{\mbb{O}}_{1, \tau}^{\dag}\right]$ is zero since 
matrix $\hat{\mbb{O}}_{1, \tau}$ is Hermitian. However, this holds true only for linearly-polarized light but not for all other types of polarization
or with a finite bandgap. 
The third term $\hat{\mbb{T}}_2(\lambda_0 \, \vert \, k, \theta_{\bf k})$ in Eq.\,\eqref{aTmexp} for a dice lattice has been calculated as

\begin{equation}
\label{T2}
\hat{\mbb{T}}_2(\lambda_0 \, \vert \, k, \theta_{\bf k}) = \frac{i\lambda_0^2v_F}{4 \sqrt{2}}\, \cos^2 \beta_0 \,
\left( k_y - k_x\tan \beta_0 \right) \,
\left[
\begin{array}{ccc}
0 & 1 & 0 \\
-1 & 0 & 1 \\
0 & -1 & 0
\end{array}
\right] =  - \frac{\lambda_0^2}{4} v_F  \cos^2 \beta_0 \,
\left( k_y - k_x \, \tan \beta_0 \right)  \, \hat{\Sigma}_y^{\,(1)} \ .
\end{equation}
Here, we would like to emphasize that if the polarization direction of the imposed radiation differs from the $x$-axis ($\beta_0\neq 0$), there exists  
an additional $k_x$ related term in Eq.\,\eqref{T2} which leads to a discontinuity for electron tunneling at the boundaries of the 
barrier region. Therefore, the boundary conditions for the components of the dressed-state wave functions in a dice lattice must be modified 
accordingly.    

\section{Dressed-State Wave Functions of Electrons in a Dice Lattice}
\label{apc}

We recall our previously derived formalism\,\cite{ourpeculiar} for the dressed-state wave functions of electrons
through finding an analytical solution at $\mbox{\boldmath$k$}= 0$ followed by seeking a general solution for the
Hamiltonian in Eq.\,\eqref{HamG} as an infinite series expansion over the complete set of 
eigenstates of $\mbox{\boldmath$k$}= 0$. 
\medskip

The obtained general solution is rather complicated and bears an explicit time dependence. However, for a dice lattice at $t=0$, we find 
the dressed-state wave function for $\gamma=\pm 1$, given by

\begin{eqnarray}
\label{new1}
&& \Psi_1^{\gamma = \pm 1} (\lambda_0, \mbox{\boldmath$k$}) = \frac{1}{2} 
\left[
\begin{array}{c}
\tau \tet{e}^{-i \Phi_1 (\lambda_0,\theta_{\bf k})} \\
\sqrt{2}\,\gamma \\
\tau \tet{e}^{i \Phi_1 (\lambda_0,\theta_{\bf k})}
\end{array}
\right] \ , \\
\nonumber 
&& \Phi_1 (\lambda_0,\theta_{\bf k}) = 2 \tan^{-1} \left[ \tau \,
\frac{J_0(\lambda_0)\,\sin \theta_{\bf k}}{f_\theta + \cos \theta_{\bf k}} 
\right] \backsimeq \tau \left[ \theta_{\bf k} - \frac{\lambda_0^2}{8} \, 
\sin^2 (2  \theta_{\bf k}) + \cdots\ \right] \ ,
\end{eqnarray}
where $J_0(x)$ is the zeroth-order Bessel function of the first kind.
The top and bottom components of the wave function in Eq.\,\eqref{new1} have equal magnitudes but differ by a phase factor only,
which is not the case for an arbitrary $\alpha-\mc{T}_3$. The phase factor $\Phi_\theta (\lambda_0)$ is not equal 
to $\theta_{\bf k} = \tan^{-1} (k_y/k_x)$ and depends on the intensity of applied radiation
($c_0$ or $\lambda_0$).
\medskip 

The remaining wave function for the flat band with $\gamma = 0$ is found to be

\begin{eqnarray}
\label{new2}
&& \Psi_1^{\gamma=0} (\lambda_0, \mbox{\boldmath$k$}) = \frac{1}{\sqrt{2}}  
\left[ 
\begin{array}{c}
 \tet{e}^{- i \Phi_0(\lambda_0,\theta_{\bf k})} \\
 0 \\
 - \tet{e}^{i \Phi_0(\lambda_0,\theta_{\bf k})}
\end{array}
\right] \ , \\
\nonumber 
&& \Phi_0(\lambda_0,\theta_{\bf k}) = \tan^{-1}\left[ 
\tau J_0(\lambda_0) \, \tan \theta_{\bf k}
\right] \backsimeq \tau \left[
\theta_{\bf k} - \frac{\lambda_0^2}{2} \, \sin (2 \theta_{\bf k}) +\cdots\
\right] \ .
\end{eqnarray}
From Eq.\,\eqref{new2}, we know that the wave function acquires only two non-zero
components with the same amplitude but different phases. Moreover, it depends on the coupling constant $\lambda_0$
but is not equal to $\theta_{\bf k}$ for the bare electron wave functions in conduction or valence bands.

\section{Boundary Conditions for Anisotropic Hamiltonian}
\label{apd}

For a pseudospin-$1$ dice lattice including a barrier region, we address the relevant boundary conditions for the case with an anisotropic Dirac cone 
and non-collinear $k_x$ and $k_{x'}$ axes. In contrast to graphene, we find that the boundary conditions for a dice lattice 
change significantly with a finite anisotropy in the energy dispersion.
\medskip

We begin from the anisotropic pseudospin-$1/2$ graphene Hamiltonian, given by  

\begin{equation}
\hat{\mc{H}}_{0}(\lambda_0, \mbox{\boldmath$k$}) = \hbar v_F \left(\hat{\Sigma}^{\left(1/2\right)}_x k_x + a_0(\lambda_0) \,
\hat{\Sigma}^{\left(1/2\right)}_y k_y \right) \ , 
\label{spinh}
\end{equation}
where $\hat{\Sigma}^{\left(1/2\right)}_x$ and $\hat{\Sigma}^{\left(1/2\right)}_y$ are the $2\times 2$ Pauli matrices, related to Eqs.\,\eqref{sig1} and \eqref{sig2}.
\medskip

Here, we consider two frames, $\{x, y\}$ and $\{x', y'\}$, where the former relates to the long-axis of an elliptical energy dispersion for dressed states of electrons while the 
latter to the normal direction of a potential barrier. 
As a result, the decomposition of a wave vector $\mbox{\boldmath$k$}$ in two frames can be written as $\{k_{x}, k_{y}\}$ or $\{k_{x'}, k_{y'}\}$, respectively, 
which are related to each other by a rotation matrix $\hat{\mbb{R}}(\beta)$, i.e.,

\begin{equation}
\label{aBetaRot}
\left[ \begin{array}{c}
k_{x} \\
k_{y}
\end{array}  \right] = \hat{\mbb{R}}(\beta)
\left[
\begin{array}{c}
k_{x'} \\
k_{y'}
\end{array} \right] \ , 
\end{equation}
where $\beta$ is the angle between two frames and 

\begin{equation}
\hat{\mbb{R}}(\beta)  = \left[
\begin{array}{cc}
\cos \beta & -\sin \beta \\
\sin \beta & \cos \beta
\end{array}
\right] \ . 
\end{equation}
In order to find the proper boundary conditions, we need transform the $k_{x,y}-$dependent Hamiltonian into $\{x',y'\}$ frame, integrate each of the equations over a small interval from $-\delta x'$ to $\delta x'$ and take the limit of $\delta x' \rightarrow 0$ afterwards.\,\cite{alphaDice, alphaKlein} 
\medskip

Let us start with the transformed dressed-state Hamiltonian for anistropic graphene within the $\{x', y'\}$ frame, given by

\begin{equation}
\hat{\mc{H}}_{0}(\lambda_0, \mbox{\boldmath$k$}) = \hbar v_F k \left[
\begin{array}{cc}
0 & \cos(\theta'_{\bf k} + \beta)- i a_0(\lambda_0) \sin(\theta'_{\bf k} + \beta) \\
\cos(\theta'_{\bf k} + \beta) + i a_0(\lambda_0) \sin(\theta'_{\bf k} + \beta) & 0
\end{array}
\right] \ , 
\label{spinh2}
\end{equation}
where $\theta_{\bf k}=\theta'_{\bf k}+\beta$, 
$\tan \theta_{\bf k} = k_y/k_x$. For the case with $a_0(\lambda_0)=1$, the transformed Hamiltonian in Eq.\,\eqref{spinh2} within the $\{x', y'\}$ becomes 

\begin{equation}
\hat{\mc{H}}_{0}(\lambda_0, \mbox{\boldmath$k$}) = \hbar v_F \left[
\begin{array}{cc}
0 & k_- \tet{e}^{- i \beta} \\
k_+ \tet{e}^{i \beta} & 0
\end{array}
\right] \ ,
\label{spinh3}
\end{equation}
where $k_\pm=k_{x'}\pm ik_{y'}$.
Since the discontinuity of $\pr/\pr x'$ due to the existence of potential barrier is associated with the $x'$ coordinate, 
by using $k_{x'}\rightarrow -i\,\pr/\pr x'$ we generalize the Hamiltonian in Eq.\,\eqref{spinh3} into

\begin{equation}
\hat{\mc{H}}_{0}(\lambda_0\, \vert \, x', k_y') = \hbar v_F \left[
\begin{array}{cc}
0 & (- i\,\pr/\pr x' - i k_{y'})\, \tet{e}^{- i \beta} \\
(- i\,\pr/\pr x'  + i k_{y'})\, \tet{e}^{i \beta} & 0
\end{array}
\right] \, , 
\end{equation}
while all the other continuous terms on both sides of the eigenvalue equation approach zero in the limit of $\delta x' \rightarrow 0$, i.e.,

\begin{eqnarray}
&& \int\limits_{-\delta x}^{\delta x} V_B \, \Theta(x) \, \varphi_j(x) \rightarrow 0 \ ,\\ 
\nonumber 
&& \int\limits_{-\delta x}^{\delta x} \varepsilon_0 \varphi_j(x) = \rightarrow 0 \ , 
\end{eqnarray}
where $\varphi_j(x)$ with $j=1,\,2$ represents one of the wave-function components.
As a result, only the terms containing $k_{x'} \rightarrow - i \, \pr/ \pr x'$ make non-zero contributions to the boundary conditions, leading to

\begin{eqnarray}
\nonumber 
&& \int\limits_{-\delta x}^{\delta x} - i \frac{\pr}{\pr x'} \, \left[ \cos \beta - i \, a_0(\lambda_0) \, \sin \beta \right]
\varphi_2 (x') = 0 \ \ \ \rightarrow \ \ \ \varphi_2(\delta x') = \varphi_2(- \delta x')\ ,\\
\label{graph}
&& \int\limits_{-\delta x}^{\delta x} - i \frac{\pr}{\pr x'} \, \left[\cos \beta + i \, a_0(\lambda_0) \, \sin \beta \right]
\varphi_1 (x') = 0 \ \ \ \rightarrow \ \ \ \varphi_1(\delta x') = \varphi_1(- \delta x')\ .
\end{eqnarray}
The obtained results are equivalent to those for the earlier considered isotropic graphene, therefore, {\it the anisotropy and the
rotation $\hat{\mbb{R}}(\beta)$ do not affect our boundary conditions}. 
\medskip 

The situation changes drastically for a dice lattice with the pseudospin-$1$ Hamiltonian. 
We once again rewrite the Hamiltonian in Eq.\,\eqref{HamG} within the $\{x',y'\}$ frame, leaving out all the continuous terms involving eigenenergy $\varepsilon_0$, piecewise 
potential $V_B \, \Theta(x)$ and constant $k_{y'}$. As a result, we only keep the terms including $- i \, \pr/ \pr x'$ and are left with

\begin{equation}
\hat{\mc{H}}_{1}^\tau (x') = \frac{\hbar v_F}{\sqrt{2}} \, \left( - i\,\frac{\pr}{\pr x'} \right) \, \left\{
\left[
\begin{array}{ccc}
0 &  \tau \cos \beta - i a_1(\lambda_0) \sin \beta  &  0 \\
0 & 0 & \tau \cos \beta - ia_1(\lambda_0) \sin \beta \\
0 & 0 & 0
\end{array}
\right] + \text{h.c} \,
\right\}\ .
\end{equation}
Correspondingly, the boundary conditions are found to be

\begin{eqnarray}
\nonumber
&&\varphi_{2}(-\delta x') = \varphi_{2}(\delta x')\ ,\\
\label{dice}
&&c_\tau^+(\lambda_0, \beta) \, \varphi_1(- \delta x') +  c_\tau^-(\lambda_0, \beta) \, \varphi_3(- \delta x') = 
c_\tau^+(\lambda_0, \beta) \, \varphi_1(\delta x') + c_\tau^-(\lambda_0, \beta) \, \varphi_3(\delta x') \ ,
\end{eqnarray}
where

\begin{equation}
c_\tau^\pm (\lambda_0, \beta) = \tau \cos \beta \pm i \, a_1(\lambda_0) \sin \beta \ .
\end{equation}
In the case with $a_1(\lambda_0)=0$ and collinear $x$ and $x'$ ($\beta = 0$), $c_\tau^\pm(\lambda_0, \beta=0) = \tau$,
and then we immediately recover the previously obtained boundary conditionsfor a dice lattice\,\cite{alphaDice} 

\begin{eqnarray}
&& \varphi_{2}(-\delta x) = \varphi_{2}(\delta x) \ , \\
\nonumber 
&& \varphi_1(- \delta x') + \varphi_3(- \delta x') = \varphi_1(\delta x') + \varphi_3(\delta x') \ .
\end{eqnarray}
For an isotropic Dirac cone but with $\beta\neq 0$ ($k_x \neq k_{x'}$), $c_\tau^\pm (\lambda_0 \rightarrow 0, \beta)  \rightarrow \tau \, \tet{e}^{\pm i \tau \beta }$, 
and the boundary conditions must be modified even for this case.  

\color{black}
\clearpage
\bibliography{AlphaKlein}

\end{document}